\documentclass[prd,twocolumn,notitlepage,preprintnumbers,nofootinbib,prl,superscriptaddress,10pt,floatfix]{revtex4-1}

\def\l@subsubsection#1#2{}
\def\l@subsubsubsection#1#2{}
\makeatother

\usepackage[normalem]{ulem}
\usepackage[utf8]{inputenc}
\usepackage{slashed}
\usepackage{enumitem}
\usepackage{graphicx,amssymb,amsmath,amsthm,amsfonts,epsfig,epsf}
\usepackage[usenames,dvipsnames,svgnames,table]{xcolor}
\usepackage{epstopdf}
\definecolor{darkred}{rgb}{0.5,0,0}
\usepackage{dcolumn}
\usepackage{latexsym}
\usepackage{rotating}
\usepackage{longtable}

\usepackage{enumerate}
\usepackage{tensor,multirow}
\usepackage{url}
\usepackage{float}
\usepackage{mathtools,slashed}
\usepackage[linktocpage]{hyperref}

\def\be{\begin{equation}}
\def\ee{\end{equation}}
\newcommand{\bea}{\begin{eqnarray}}
\newcommand{\eea}{\end{eqnarray}}

\def\ba{\begin{align}}
\def\ea{\end{align}}

\allowdisplaybreaks

\begin{document}

\title{{\bf Precessing binary black holes as engines of electromagnetic helicity}}

\author{Nicolas Sanchis-Gual}
\affiliation{Departament d'Astronom\'{\i}a i Astrof\'{\i}sica, Universitat de Val\`encia,
		Dr. Moliner 50, 46100, Burjassot (Val\`encia), Spain}	
	\affiliation{Departamento  de  Matem\'{a}tica  da  Universidade  de  Aveiro  and  Centre  for  Research  and  Development in  Mathematics  and  Applications  (CIDMA),  Campus  de  Santiago,  3810-183  Aveiro,  Portugal} 
\author{Adrian del Rio}
\affiliation{Departamento de F\'isica Te\'orica and IFIC, Universitat de Valencia-CSIC. Dr. Moliner 50, 46100, Burjassot (Valencia), Spain.}

\begin{abstract}
We show that  binary black hole mergers  with precessing evolution can potentially excite photons from the quantum vacuum in such a way that total helicity is not preserved in the process. Helicity violation is allowed by quantum fluctuations that spoil the  electric-magnetic duality symmetry of the classical Maxwell theory without charges. We show here that precessing binary black hole systems in astrophysics generate a flux of circularly polarized gravitational waves which, in turn, provides the required helical background  that triggers this quantum effect. Solving the fully non-linear Einstein's equations with numerical relativity  we explore  the parameter space of  binary systems and extract the detailed dependence of the quantum effect with the spins of the two black holes. We also introduce a set of diagrammatic techniques that allows us to predict when a binary black hole merger can or cannot emit  circularly polarized gravitational radiation, based on  mirror-symmetry considerations. This framework allows to understand and to interpret correctly the numerical results, and to predict the outcomes in potentially interesting astrophysical systems. 
\end{abstract}

\date{\today}

\maketitle

%%%%%%%%%%%%%%%%%%%%%%%%%%%%%
\section{Introduction}
%%%%%%%%%%%%%%%%%%%%%%%%%%%%%

A dynamical spacetime can excite the quantum modes of the electromagnetic field and can produce as a result  photons out of the quantum vacuum \cite{birrell-davies, parker-toms}.
Well-known examples of this effect  were explored long ago in cosmological backgrounds  \cite{Parker68} and in the gravitational collapse of stars  \cite{Hawking75}. 
The particles created by the spacetime are entangled and in particular their physical properties  respect  the symmetries of the background.  
For instance, if the spacetime is spatially homogeneous, as is typical in cosmology,  particles are produced in pairs and with opposite linear momentum. This is because of the invariance of the field modes or vacuum state under spatial translations. Similarly,  the spherical symmetry of the Schwarzschild metric in a gravitational collapse requires that the Hawking pairs have opposite angular momentum.  In other words, the symmetries of the background impose constraints on the particles created. If, on the other hand, the background spacetime does not possess these symmetries, then the particles created may not be subject to such limitations. To give an example, for the gravitational collapse of a rotating star, where spherical symmetry is lost, the spacetime dynamics can induce a net angular momentum in the flux of particles created, particle pairs are not necessarily created with opposite angular momenta \cite{Page1976B}.

In addition to the symmetries of the background, there are intrinsic symmetries of the quantum field that must be preserved during the process of particle creation. For instance, the electromagnetic theory must be gauge invariant, and if the electromagnetic field is coupled to fermion fields, this symmetry requires  the conservation of the electric charge in any  process. Interestingly, in some particular cases the background spacetime can induce fundamental violations of classical internal symmetries in the quantum theory. An example of this is the electric-magnetic duality symmetry of the  source-free Maxwell theory. In the classical theory this symmetry guarantees that the circularly polarized state of  electromagnetic waves remains constant during their propagation. Then, one could naively expect  that, in any dynamical gravitational field, photons should be created in pairs of opposite helicity, so as to keep the same circular polarization state of the vacuum. However, it was found that this symmetry fails to survive the quantization in a gravitational field \cite{AdRNS2017a,AdRNS2017b,AdRNS2018a, AdRNS2018b}. As a result, the net helicity {need} not be conserved, and photons {are expected to} be created without having to satisfy this constraint, provided the background spacetime is helical. 

{Given a fixed spacetime background that evolves between two asymptotically stationary configurations, a detailed study of  how many  photons  are created in each helicity sector from this anomaly, as well as the  frequency and angular spectrum, requires an explicit calculation of the Bogoliubov coefficients that relate ``in'' and ``out'' vacuum states \cite{birrell-davies}. However, except for few well-known examples, this calculation is  inaccessible with current theoretical techniques. Despite this, it is still possible to determine the average total amount of right-handed minus left-handed photons created. This quantity {\it is accessible} from the vacuum expectation value of the operator $Q_5$ that represents the classical Noether charge in the quantum theory. Indeed, the quantum anomaly indicates that the change in time of this expectation value is independent of the choice of quantum state and, furthermore, it only depends on the background geometry  as:}
\begin{align}
\Delta\hat Q_{5}&\equiv\langle \hat Q_5(t_2)\rangle - \langle\hat Q_5(t_1)\rangle \nonumber\\&=\frac{-{\hbar}}{96\pi^2} \int_{[t_1,t_2]\times \Sigma} d^4x \sqrt{-g} R_{abcd}{^*R}^{abcd}, \label{gravcp}
\end{align}
where $R^a_{\hspace{0.15cm}bcd}$ denotes the Riemann tensor of the  spacetime. {The quotient $\Delta\hat Q_{5}/\hbar$ is  the  net average number difference between positive-helicity photons (or right-handed) and negative-helicity photons (or left-handed) created by the gravitational dynamics (integrated over all possible frequencies and momenta). Since this is fully determined by the spacetime geometry, it can be evaluated very easily with usual techniques in General Relativity.} 
In compact manifolds without boundary the right-hand side (RHS) of Eq.~(\ref{gravcp}) is a topological invariant, called the Chern-Pontryagin scalar. In General Relativity and astrophysics, 4-dimensional spacetime manifolds of physical interest are neither compact nor boundaryless, and the Chern-Pontryagin does contain information about the geometry (i.e. about the gravitational field). Roughly speaking, it measures the helical nature of the spacetime. This is, the degree of gravitational chirality.

In a previous paper we proved that this quantum anomaly is produced whenever the spacetime background admits a flux of {net} circularly polarized gravitational radiation \cite{dRSGMAFNS, dR21}. More precisely:
\begin{eqnarray}
\langle\hat Q_5(\mathcal{J}^+)\rangle - \langle\hat Q_5(\mathcal{J}^-)\rangle = \text{(...)}+\hspace{3cm}\label{GWcircular}\\
{\hbar}\int_0^{\infty} \frac{d\omega \omega^3}{24\pi^3} \sum_{\ell m} \bigl[| h_+^{\ell m}(\omega)-ih^{\ell m}_{\times}(\omega)|^2\nonumber\\
-|h_+^{\ell m}(\omega) + ih^{\ell m}_{\times}(\omega)|^2 \bigl] \, , \nonumber
\end{eqnarray}
where $h_+$, $h_{\times}$ denote the two  linear polarization modes of gravitational waves that reach future null infinity, emitted by an arbitrary isolated gravitational source that is stationary at both past and future timelike infinities. These modes are characterized by the frequency $\omega$, and angular momentum $\ell, m$. The contribution denoted by dots corresponds to the flux of chiral gravitational flux falling through the black hole (BH) horizon. The explicit expression is tedious but will not be relevant in our discussion. The physical picture is simple: a non-trivial gravitational field can create a difference in the number of right- and left-handed circularly polarized photons from the quantum vacuum. The more right(left)-handed gravitational radiation is emitted by a system, the more right(left)-handed electromagnetic modes will be excited.

In this paper we  examine in great detail which spacetime backgrounds in astrophysics can generate such gravitational wave flux.  Using symmetry arguments and some diagrams we will be able to predict that precessing binary BH systems can potentially trigger this quantum effect. We will prove this rigorously solving the fully non-linear Einstein's equations using standard techniques in numerical relativity, and explore the dependence with the {relative masses} and spins of the BHs. {Notice on the other hand that the net difference of positive and negative photons (\ref{gravcp})} will be insensitive to the total mass of the system, {since the integral on the RHS is adimensional and one can always rescale the coordinates by this mass.} 

{Along this paper we work in geometric units $G=c=1$. The present paper is a detailed exposition of the numerical results presented in \cite{dRSGMAFNS}, where the main results were communicated.}

%%%%%%%%%%%%%%%%%%%%%%%%%%%%%
\section{Binary diagrams and mirror symmetry}
%%%%%%%%%%%%%%%%%%%%%%%%%%%%%

Although it may seem a trivial question, it is actually difficult to find examples of physically interesting gravitational fields that make (\ref{gravcp}) non-zero. In fact, one can prove that all stationary, asymptotically flat solutions of Einstein's equations lead to a vanishing result \cite{dR21}. As a consequence, one needs dynamical gravitational fields in the fully non-linear regime, and, in turn, this requires the use of numerical relativity.

Unfortunately, solving Einstein's equations numerically is a computationally expensive task. To study this question efficiently, 
it is necessary first to have some guidance. If one restricts to binary BH systems in astrophysics, it is possible to infer which family of solutions can be expected to produce non-trivial results using just symmetry arguments. The key idea is to notice that (\ref{gravcp}) is a pseudo-scalar. As a result, any binary system that is invariant under a mirror transformation with respect to, at least, one coordinate plane, will make this integral equal to zero. The goal then is to look for systems with no mirror symmetries.

Let us make this idea more precise. {Consider a 3+1 foliation of the spacetime manifold $M=I\times \Sigma$.
In 3+1 numerical relativity Einstein's equations are solved  with 3-dimensional euclidean grids, so we will  restrict to  spatial slices with trivial  topology, $\Sigma\simeq\mathbb R^3$}  \footnote{{
For spacetimes involving black holes a convenient 3+1 foliation is engineered to  bypass the curvature singularities, in such a way that they remain in the asymptotic future of $\Sigma$ and Eintein's equations are well-posed. The spatial slices $\Sigma$ are therefore not ``pierced'' by singularities, they remain smooth~\cite{hannam2007geometry,alcubierre2008introduction}. An illustrative example is given by the usual Penrose diagram for a spherically symmetric collapse. It is possible to foliate the spacetime by  spacelike hypersurfaces $\Sigma\simeq \mathbb R^3$, and they  only intersect the curvature singularity for $t\to \infty$.}}
The different binary BH systems are uniquely represented by a 4-dimensional metric $g_{ab}$, that is solution of Einstein's equations. For each of them we can calculate the time-dependent quantity $F[g_{ab}](t)=\int_{\Sigma}d^3x\sqrt{-g}R_{abcd}{^*R}^{abcd}$. We can think of this as a quantity that keeps track of the chirality of the gravitational field as a function of time.  As  a pseudo-scalar it flips sign under a reflection $I$ of the metric (improper rotation) and remains invariant under a proper rotation $R$ of the metric, namely $F[(R\circ I)g_{ab}]=-F[g_{ab}]$. If the metric of a binary system  is invariant under a mirror transformation with respect to some coordinate plane, then one also has $F[(R\circ I)g_{ab}]=F[g_{ab}]$, and therefore $F[g_{ab}]=0$ in these cases.
\begin{figure}[t!]
\centering
\includegraphics[width=1.0\linewidth]{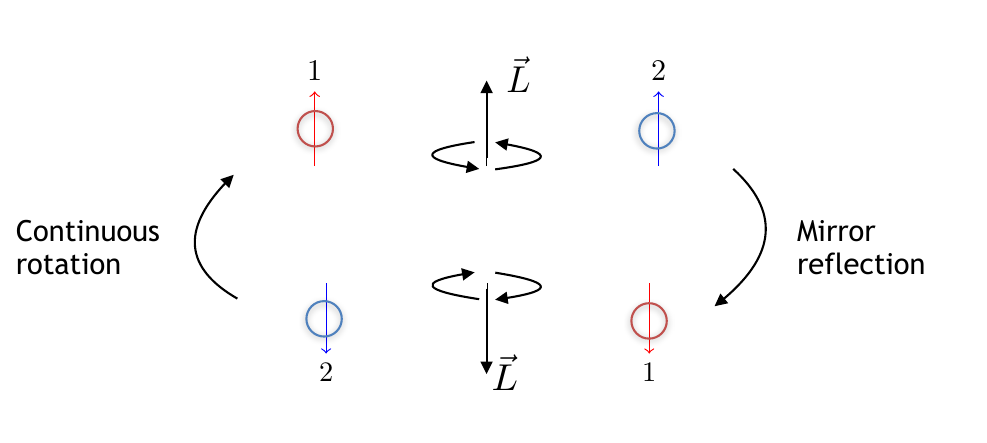}
\caption{Example of a binary BH system that is expected to yield a zero value of the Chern-Pontryagin (\ref{gravcp}). The picture represents one instant of time of a non-precessing binary system with orbital angular momentum $\vec L$. The arrows 1 and 2 denote the individual spin vectors of each BH, and they keep aligned with $\vec L$ the whole evolution. The existence of a mirror symmetry in the metric produces $F[g_{ab}](t)=\int_{\Sigma}d^3x\sqrt{-g}R_{abcd}{^*R}^{abcd}=0$ for any $t$. Numerical simulations confirm this theoretical prediction.}
\label{diagram}
\end{figure}

\begin{figure}[t!]
\centering
\includegraphics[width=1.0\linewidth]{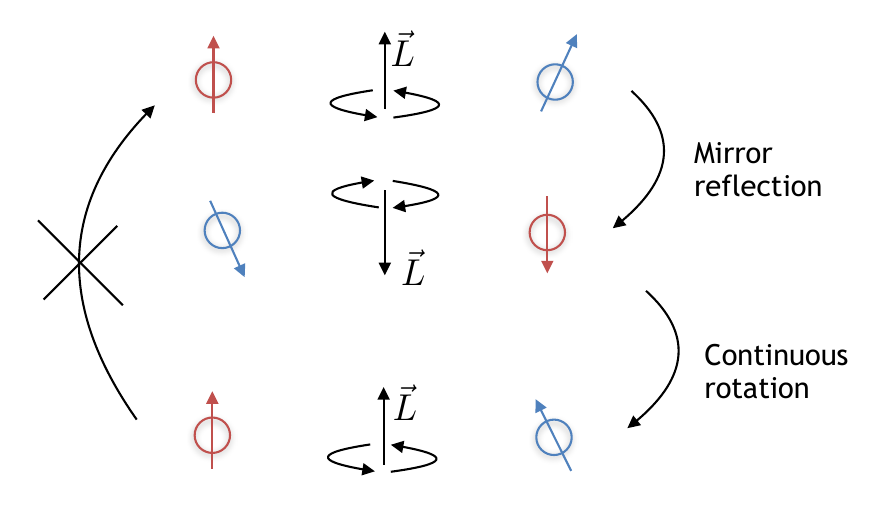}
\caption{Example of a binary BH system that is expected to yield a non-zero value of the Chern-Pontryagin (\ref{gravcp}). The picture represents one instant of time of the evolution of the binary, which has orbital angular momentum $\vec L$.  The red and blue arrows  denote the individual spin vectors of each BH, and are misaligned with $\vec L$. This produces the system to precess in time. The lack of a mirror symmetry in the metric can potentially yield $F[g_{ab}](t)=\int_{\Sigma}d^3x\sqrt{-g}R_{abcd}{^*R}^{abcd}\neq 0$ at each $t$.  Numerical simulations confirm these theoretical expectations. In particular, the merger produces a flux of circularly polarized gravitational waves, as predicted in equation (\ref{GWcircular}).}
\label{diagram2}
\end{figure}

\begin{figure*}[t!]
\centering
\includegraphics[width=0.501\linewidth]{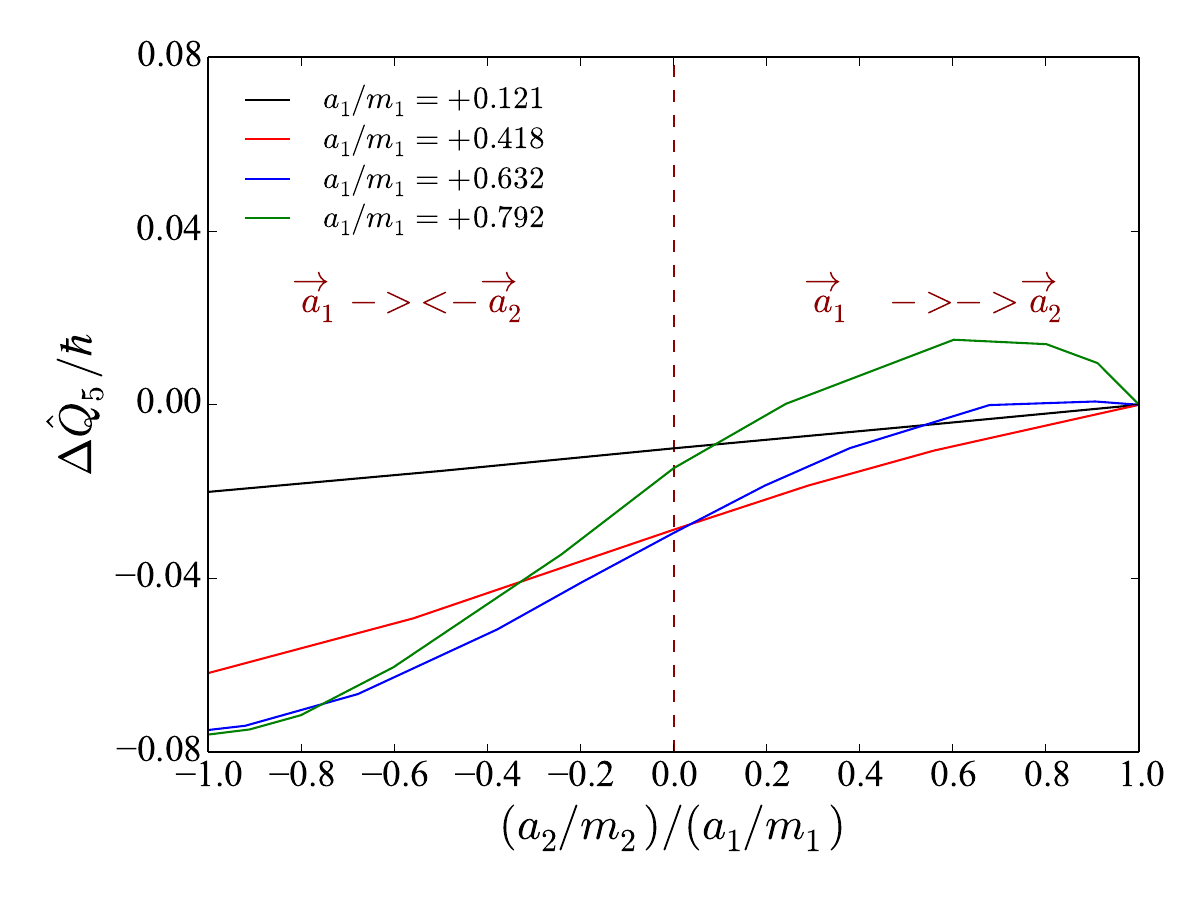}\hspace{-0.25cm}\includegraphics[width=0.501\linewidth]{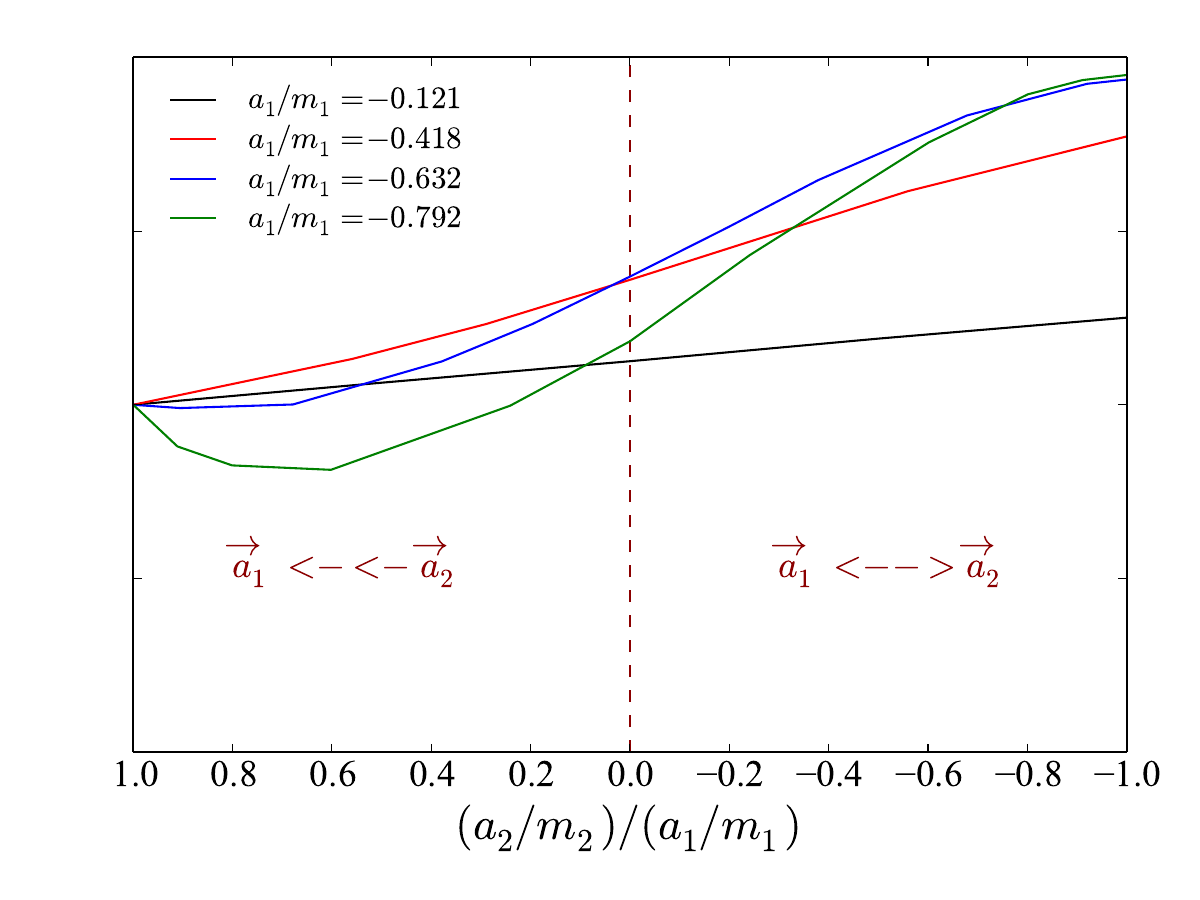}\caption{Value of $\Delta \hat Q_5/\hbar$ calculated from (\ref{EB}) in head-on collisions as a function of the spin ratio $a_2/a_1$ for four different fixed values of $a_1$. {As $a_2$ evolves from negative to positive values and viceversa, the binary system transitions between two relative spin orientations, indicated in each figure with two  vectors on the real line}. Notice how $\Delta \hat Q_5/\hbar$ flips sign {when switching} between the two figures,  as expected from the mirror transformation underlying the figures.}
\label{figHON}
\end{figure*}

To give an illustrative example, consider a binary BH system in which the two spins are parallel to the orbital angular momentum, as in Fig. \ref{diagram}. These systems are non-precessing, the orientation of the orbital angular momentum is constant (roughly speaking, the two BHs remain in a plane all the time). As a first approximation, we can assume that the gravitational field of the binary is equivalent to the gravitational fields of the two individual Kerr BHs (i.e. we ignore the non-linearities associated to the mutual interaction). BHs are rigid compact objects, in the sense that tidal love numbers are zero or very small, so this approximation should work  well. In this approximation, the entire spacetime geometry will be determined by the two masses and the two spins, because of the no-hair theorem. 
A simple analysis using symmetry arguments  allows us to infer which binary systems can produce circularly polarized gravitational waves, i.e. if $F[g_{ab}](t)\neq 0$. 
First of all, take the system in a fixed instant of time, like in the upper figure of Fig. \ref{diagram}.  Now perform a mirror transformation with respect to the {coordinate plane normal to the separation}
%plane of separation 
between the two objects. The result is shown in the lower part of Fig \ref{diagram}. Notice that the spins are pseudo-vectors, so one has to reverse sign under this transformation. Then, it is easy to see that we can find a continuous rotation in 3-space that returns the system back to the original configuration of masses and spins.
This simple example shows that non-precessing binary BH systems have a  mirror symmetry at any given time.  Because $F[g_{ab}](t)$   is a pseudo-scalar, it  flips sign under mirror reflection. So at each instant of time we must necessarily have $F[g_{ab}]=0$. In particular it also applies to non-spinning binary BHs, even in the unequal-mass case.

Most interestingly, the contra-positive of this statement tells us that for a spacetime to have a non-vanishing Chern-Pontryagin,  it is required that the individual BH spins must be misaligned with the orbital angular momentum. In other words, precessing binary BH systems can potentially lead to non-vanishing values of (\ref{gravcp}) and (\ref{GWcircular})\footnote{Not all precessing BH system will lead to a non-zero effect as we will see in the next secion.}. See Fig. \ref{diagram2} for an example of this.  The mirror-symmetry arguments introduced in this section  turn out to be really helpful in understanding the  outcomes of numerical simulations.

%%%%%%%%%%%%%%%%%%%%%%%%%%%%%
\section{Numerical results for precessing BHs}
%%%%%%%%%%%%%%%%%%%%%%%%%%%%%
\begin{table*}%[h!]
\begin{center}
\begin{tabular}{|c|c|c|c|c|}
\hline
Configuration&Initial spin orientation $(x, y, z)$&$|a_{i}/m_{i}|$&Total ADM mass  & $\Delta\hat Q_5/\hbar$  \\
\hline
%2.5  &  0.3993&    -1.521&2.549e-01\\
S1 &$(\leftarrow,0,\uparrow),\, (\rightarrow,0,\uparrow)$&0.312& 1.03&0.040\\
S2 &$(\leftarrow,0,\uparrow),\, (\rightarrow,0,\uparrow)$&0.520& 1.12&-0.039 \\
S3 &$(\leftarrow,0,\uparrow),\, (\rightarrow,0,\uparrow)$&0.630& 1.22&0.064 \\
%\hline
S4 &$(\rightarrow,0,\uparrow),\, (\rightarrow,0,\uparrow)$&0.520& 1.12&1.09$\times10^{-09}$\\
S5 &$(0,0,\uparrow),\, (0,0,\uparrow)$&0.630& 1.22&3.42$\times10^{-11}$ \\

\hline
X1 &$(\leftarrow,0,0),\, (\rightarrow,0,0)$&0.312& 1.03&-0.051\\
X2 &$(\leftarrow,0,0),\, (\rightarrow,0,0)$&0.520& 1.12&0.105\\
X3 &$(\leftarrow,0,0),\, (\rightarrow,0,0)$&0.630& 1.22&0.086\\
%\hline
X4 &$(\rightarrow,0,0),\, (\rightarrow,0,0)$&0.630& 1.22&1.15$\times10^{-09}$ \\
\hline

\end{tabular}
\end{center}
\caption{Value of the  Chern-Pontryagin $\Delta\hat Q_5/\hbar$ computed from Eq.~(\ref{EB}) using numerical-relativity simulations of  binary BH mergers of equal mass and spin magnitude with orbital evolution. The S configurations correspond to  binary systems where the initial BH spins  have two non-vanishing cartesian components (of the same magnitude) in our coordinate system, as indicated in the second column, while the X configurations are binary BHs where the initial spins are aligned in the $x$ direction. The spin orientations vary  cyclically during the entire evolution. Roughly, they return to the same relative orientation after one orbital period.  
The results for $\Delta\hat Q_5/\hbar$ confirm the theoretical predictions described in Sec. II using symmetry arguments. In particular, those binary BHs with a configuration of spins with mirror symmetry produce a zero value of $\Delta\hat Q_5/\hbar$ (compatible with numerical inacuracies, {see footnote 3}).  } 

\label{tabhdbbh} 
\end{table*}

In the previous section we argued that precessing binary BH systems are the relevant configurations to explore the quantum effect of equation (\ref{gravcp}).
In this section we confirm these theoretical expectations and extract the dependence of this quantity with the parameters of the binary. 

To achieve this we perform numerical simulations using the 3+1 Numerical Relativity code {\tt Einstein Toolkit}~\cite{toolkit2012open,loffler2012einstein}, and the {\tt McLachlan} thorn~\cite{brown2009turduckening,reisswig2011gravitational} for the spacetime evolution. We solve Einstein's equations for head-on, eccentric, and quasi-circular BBH mergers, taking the component masses and initial linear momentum from~\cite{tichy2004quasiequilibrium}. To compute Eq.~(\ref{gravcp}) we notice that 
\bea
\Delta\hat Q_{5} = -\frac{\hbar}{6\pi^2} \int_{t_1}^{t_2}dt \alpha\int_{\Sigma_t}d\Sigma_t\sqrt{h}E_{ij}B^{ij}, \label{EB}
\eea
where $\Sigma_t$ is a spacelike hypersurface in our 3+1 spacetime foliation, $\alpha$ is the lapse function, $h_{ij}$ is the induced metric, and $E_{ij}$, $B_{ij}$ are the electric and magnetic components of the Weyl tensor on $\Sigma_t$. We compute this by modifying the {\tt Antenna} thorn~\cite{baker2002lazarus,campanelli2006lazarus} and the initial data are obtained using the {\tt TwoPunctures} thorn~\cite{ansorg2004single}.  {As we will see later, the result of (\ref{EB}) will be dominated by the merger stage, so the specific choice of initial  momenta and initial radial separation of the two black holes will not play a significative role}. 
Our initial numerical grid is a superposition of two individual grids centered at the initial positions of the BHs. We make use of the {\tt PunctureTracker} thorn, that tracks the location of each BH puncture during the evolution. Each individual grid has 9 refinement levels with $\lbrace(320, 160, 80, 40, 20, 5, 2.5, 1.25, 0.625)$, $(4, 2, 1, 0.5, 0.25, 0.125, 0.0625, 0.03125,0.015625)\rbrace$. The first set of numbers indicates the {size of the} spatial domain of each level and the second set indicates the resolution. No symmetries are imposed on the numerical grids, therefore we have: $ x_{\rm{min}}= y_{\rm{min}}= z_{\rm{min}}=-320$ and $ x_{\rm{max}}= y_{\rm{max}}= z_{\rm{max}}=320$. We also use Carpet Adaptative Mesh Refinement for the Cactus framework~\cite{CarpetCode:web} within the {\tt Einstein Toolkit} infrastructure.  

\subsection{Head-on collisions}
Head-on collisions provide the simplest setting to study the dependence of the Chern-Pontryagin with the spins in a binary system. In contrast to orbital mergers,  the relative spin configuration remains roughly constant during the entire evolution, so that it is relatively easy to understand and interpret correctly the  numerical results in terms of the framework described in Sec. II. The numerical exploration of head-on collisions can be particularly useful if we let the  two individual BH spins be aligned with the velocity axis, say in the $x$ direction of our cartesian coordinate system, as ~$(\rightarrow,0,0),\, (\leftarrow,0,0)$ and ~$(\leftarrow,0,0),\, (\rightarrow,0,0)$. These two spin configurations are related by a mirror transformation, and both are expected to make (\ref{gravcp}) different from zero using the symmetry arguments of Sec. II. 

To explore the impact of the spin magnitude 
on (\ref{gravcp}), we evolve a series of  head-on collisions with these two spin configurations,  fixing  the spin magnitude of one of the BHs, $a_{1}/m_1$, and varying the other one in the range $a_{2}/m_2\in( -a_{1}/m_1,a_{1}/m_1)$. The initial separation of the two BHs is not expected to play any important role in this problem so in all cases we fix $D=11$ in code units. 
The results of 4 representative cases are summarized in Figure~\ref{figHON}, where we plot the values of the Chern-Pontryagin from Eq.~(\ref{EB}) as a function of the ratio $(a_{2}/m_2)/(a_{1}/m_1)$ for four values of $a_{1}/m_1$. We conclude that the Chern-Pontryagin (\ref{EB}) reaches its maximum (in absolute value) when the two BHs have spins with equal magnitudes but opposite direction, while the $a_{1}=a_{2}$ (with $m_1=m_2$) configuration gives a zero contribution. In addition, the right panel of Fig.~\ref{figHON} shows that flipping the sign  of $a_1$ only results in an overall change of sign in the Chern-Pontryagin, keeping the same magnitude. All these results confirm the validity of the analysis of mirror symmetry described in Sec. II above. It is worth noticing that even in the collision of a Kerr and of a Schwarzschild BHs, the resulting effect is non-zero (see Fig.~\ref{figHON}). 

\begin{figure*}[t!]
\centering
\includegraphics[width=0.329\linewidth]{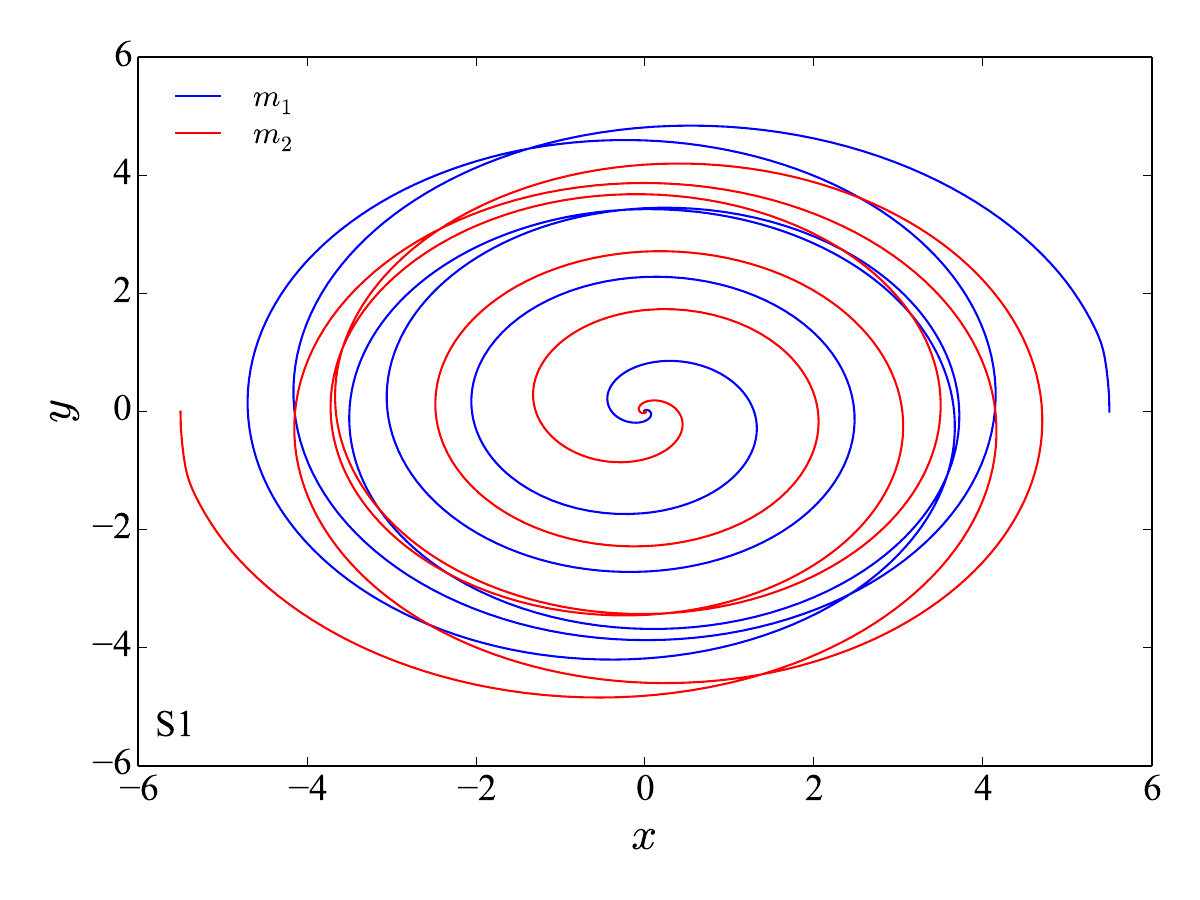}\includegraphics[width=0.329\linewidth]{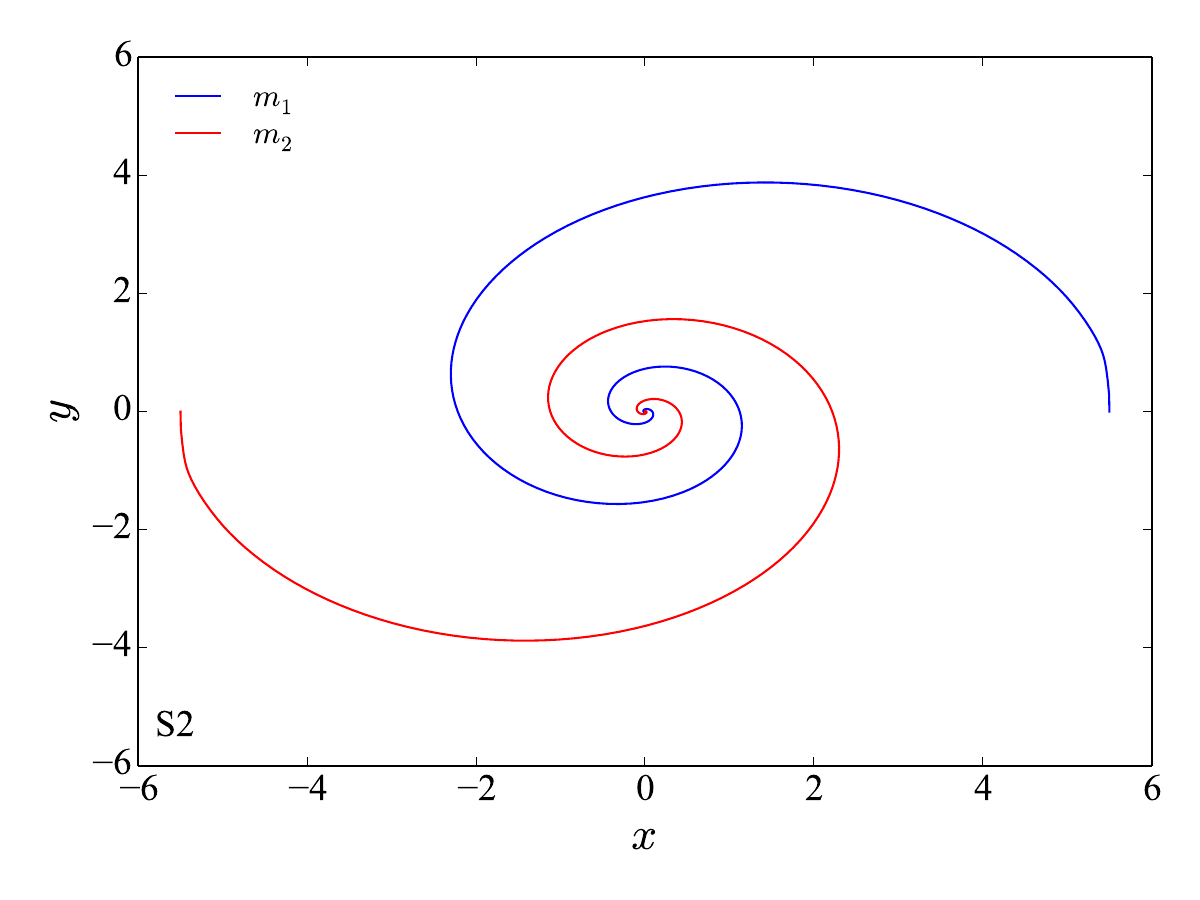}
\includegraphics[width=0.329\linewidth]{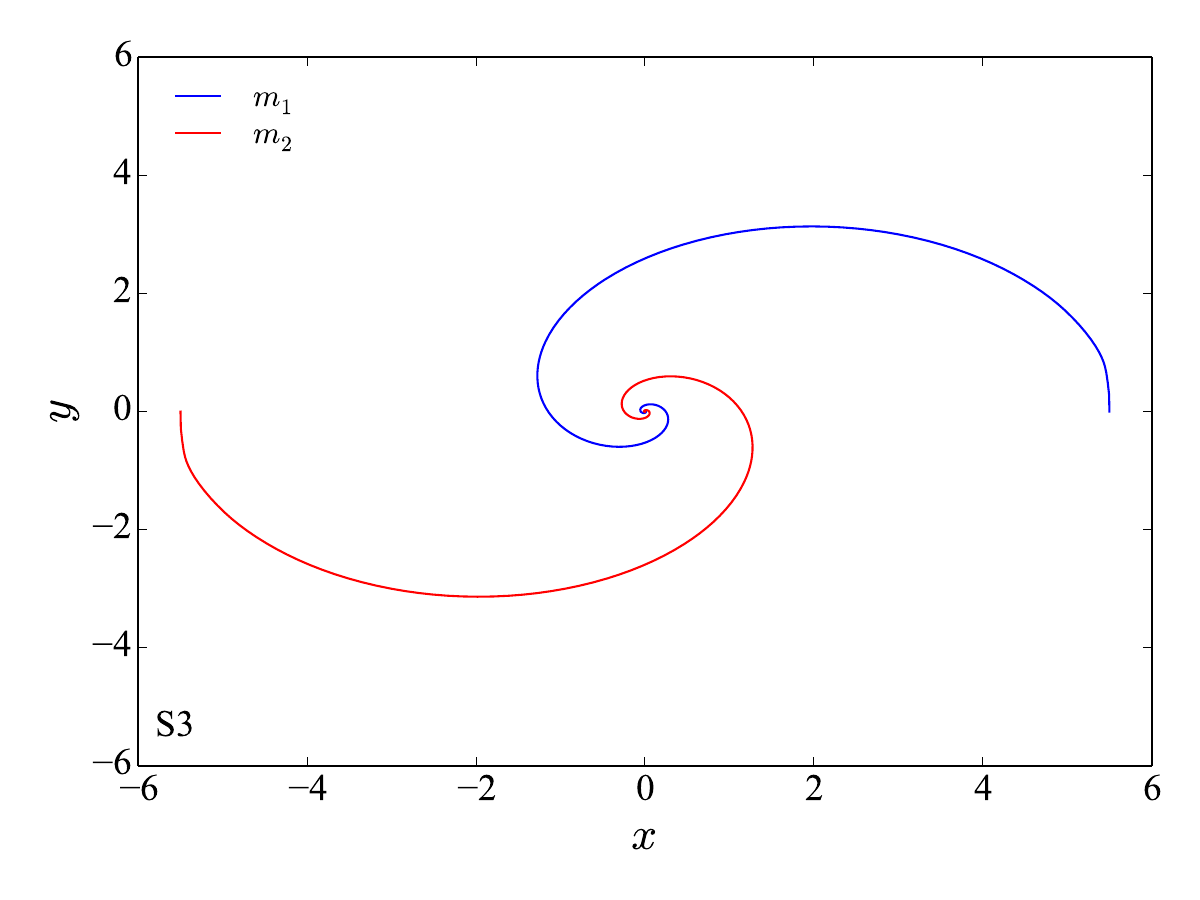}\\\includegraphics[width=0.329\linewidth]{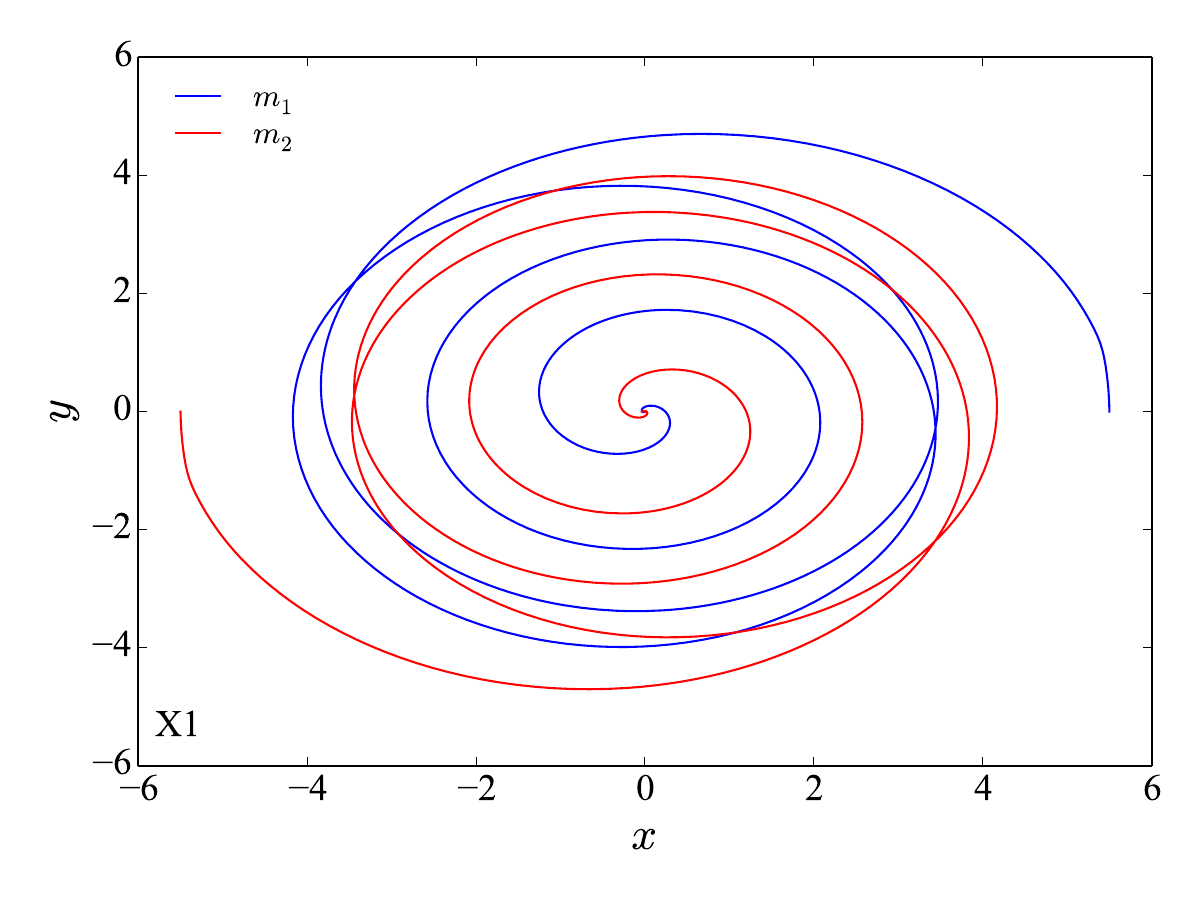}
\includegraphics[width=0.329\linewidth]{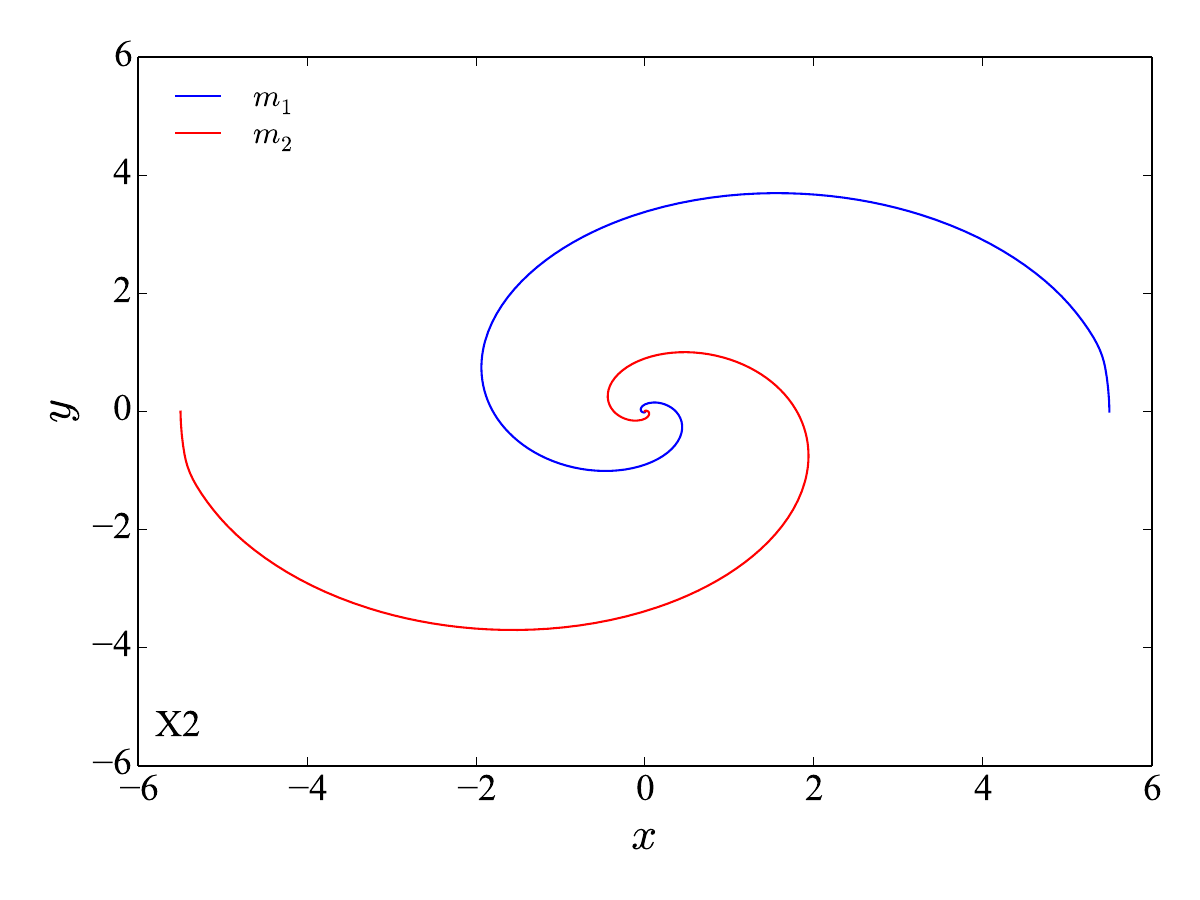}
\includegraphics[width=0.329\linewidth]{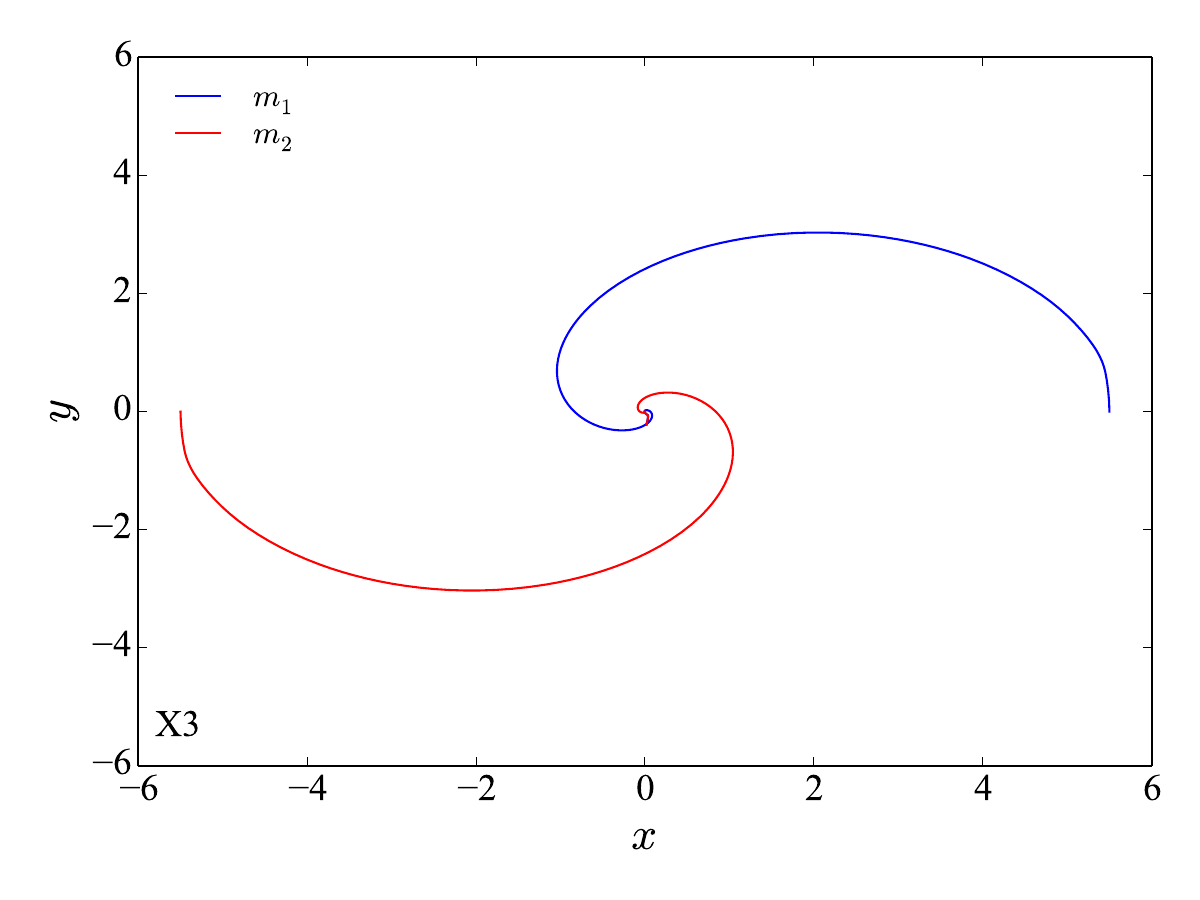}\\\includegraphics[width=0.329\linewidth]{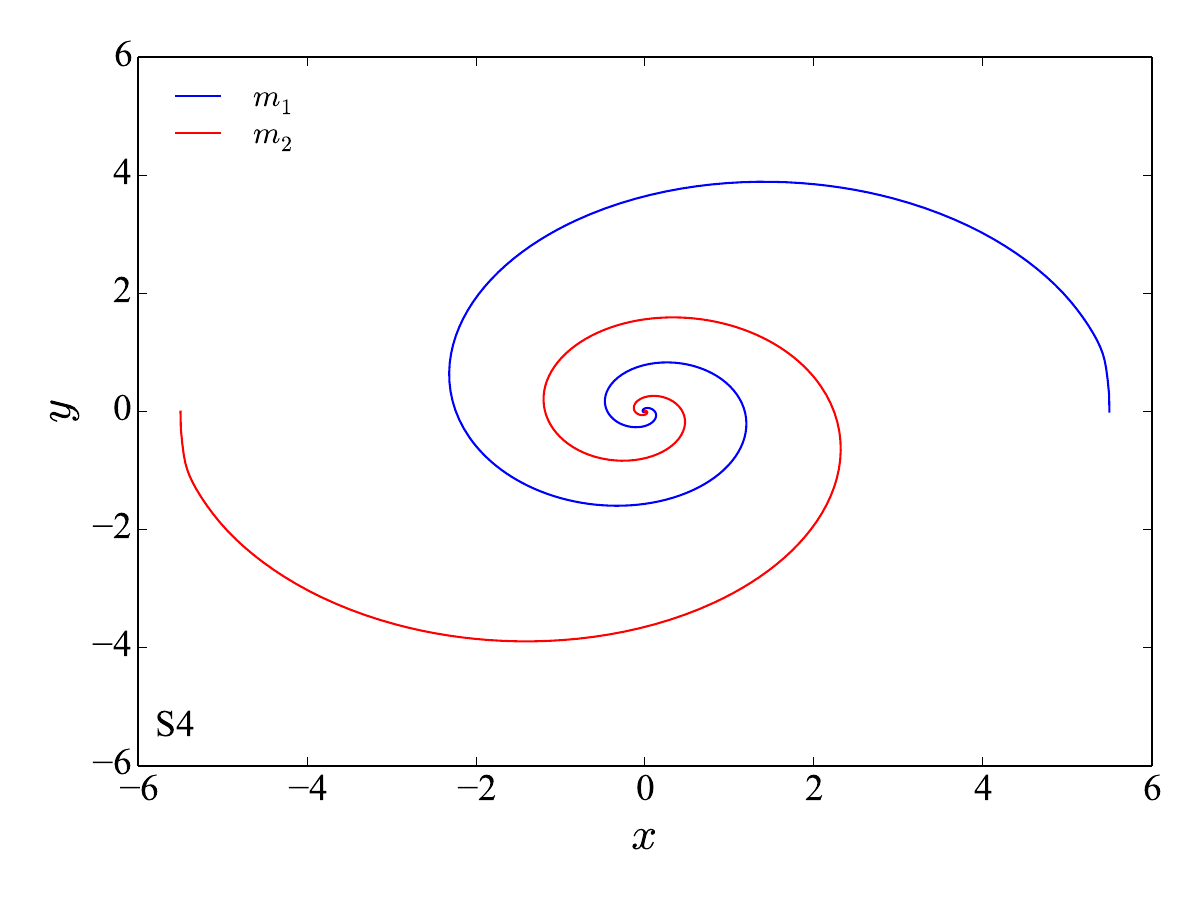}
\includegraphics[width=0.329\linewidth]{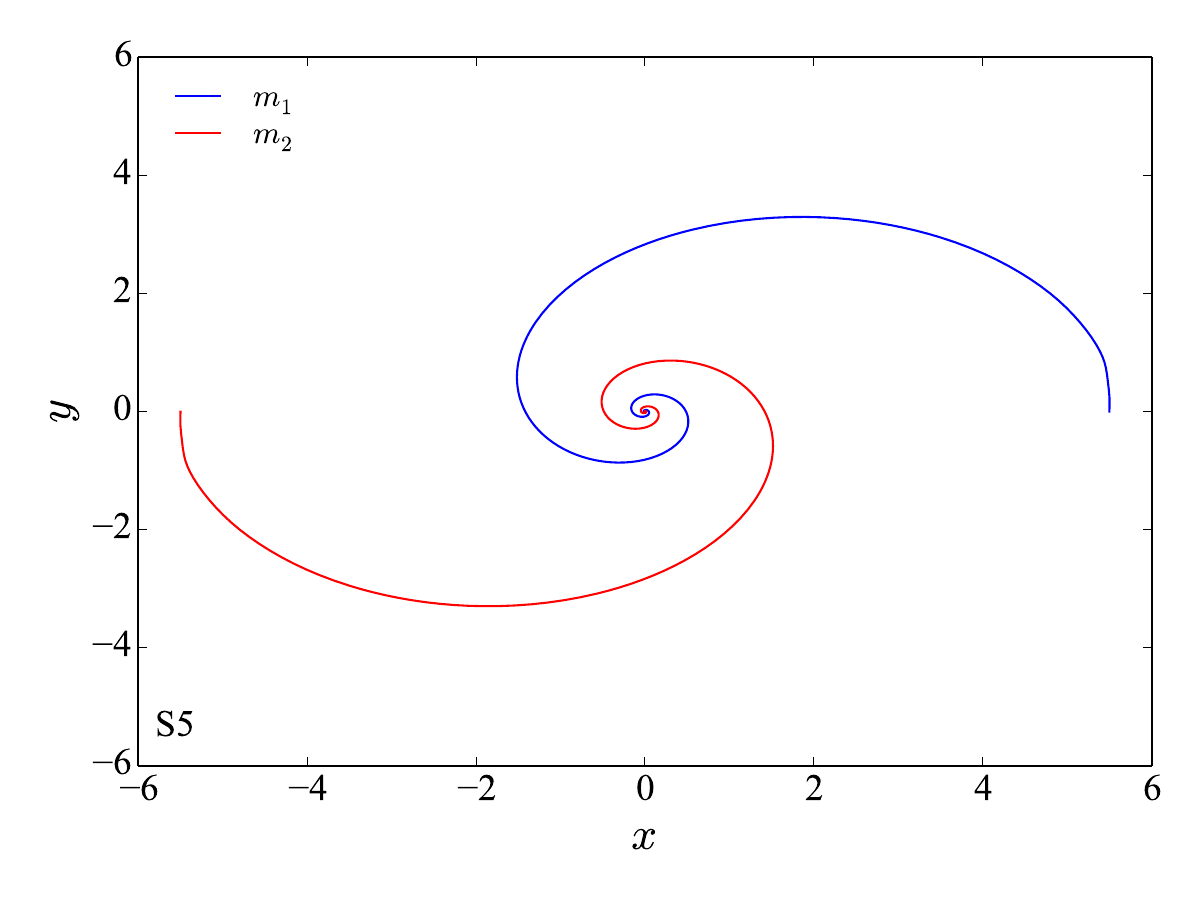}\includegraphics[width=0.329\linewidth]{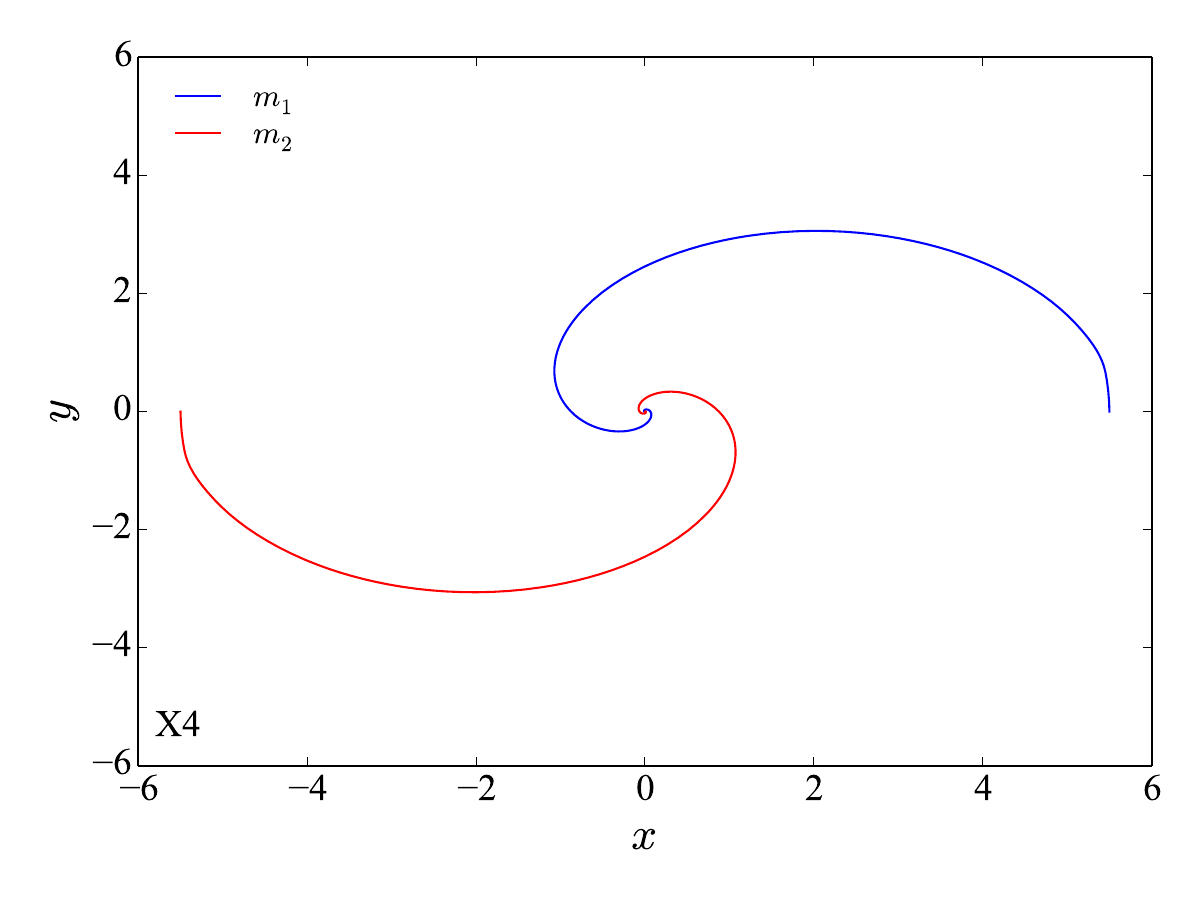}
\caption{Trajectories of the BHs {projected on}  the equatorial $xy$ ($z=0$) plane for the different configurations described in Table~\ref{tabhdbbh}. }
\label{figTraj}
\end{figure*}

\begin{figure*}[t!]
\centering
\includegraphics[width=0.329\linewidth]{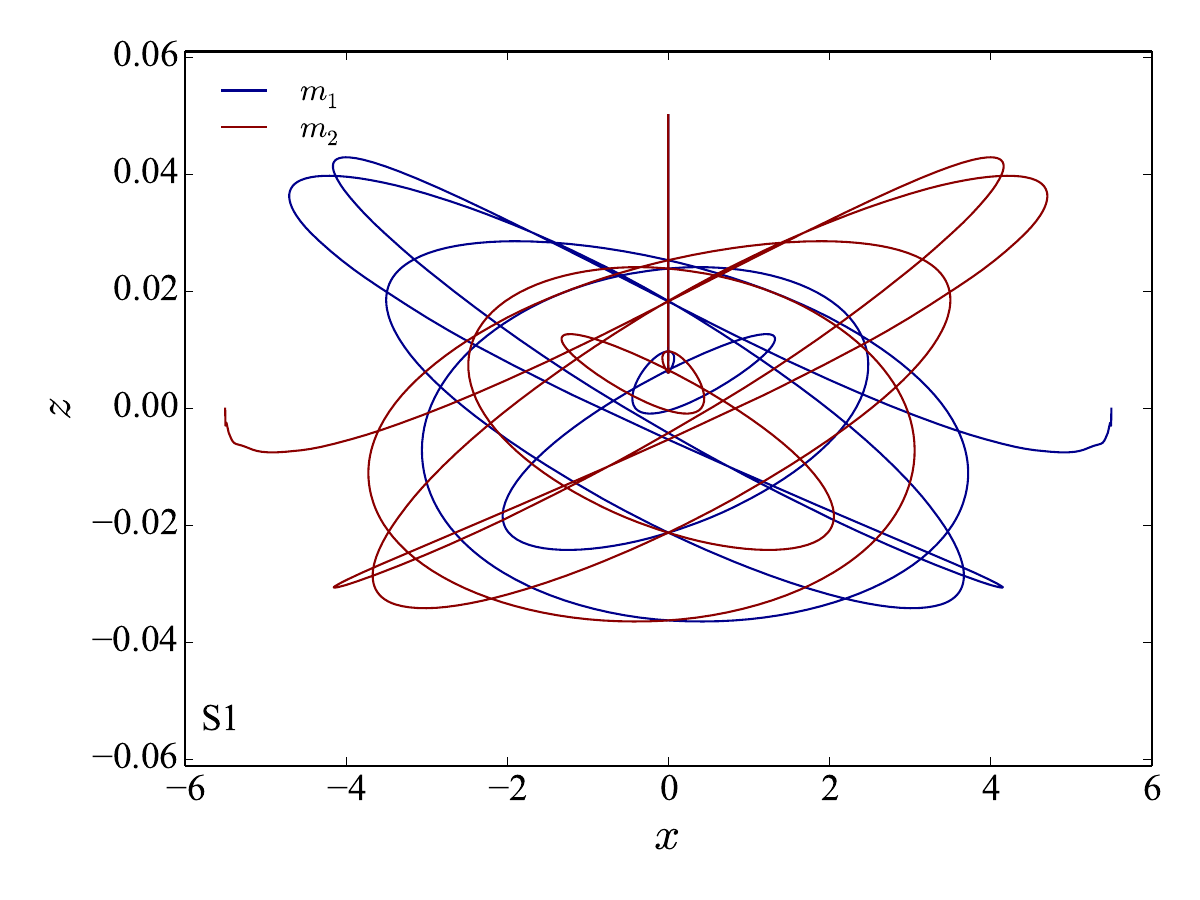}\includegraphics[width=0.329\linewidth]{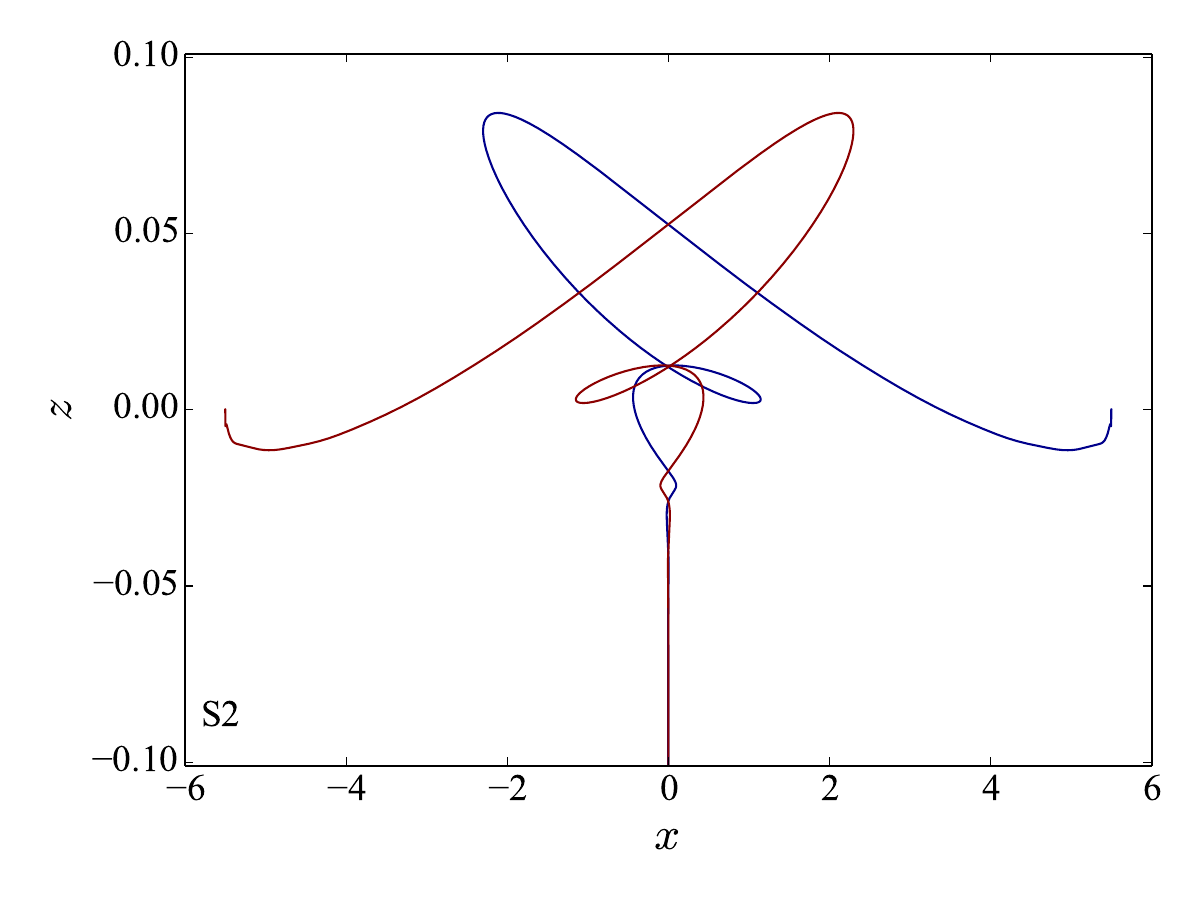}
\includegraphics[width=0.329\linewidth]{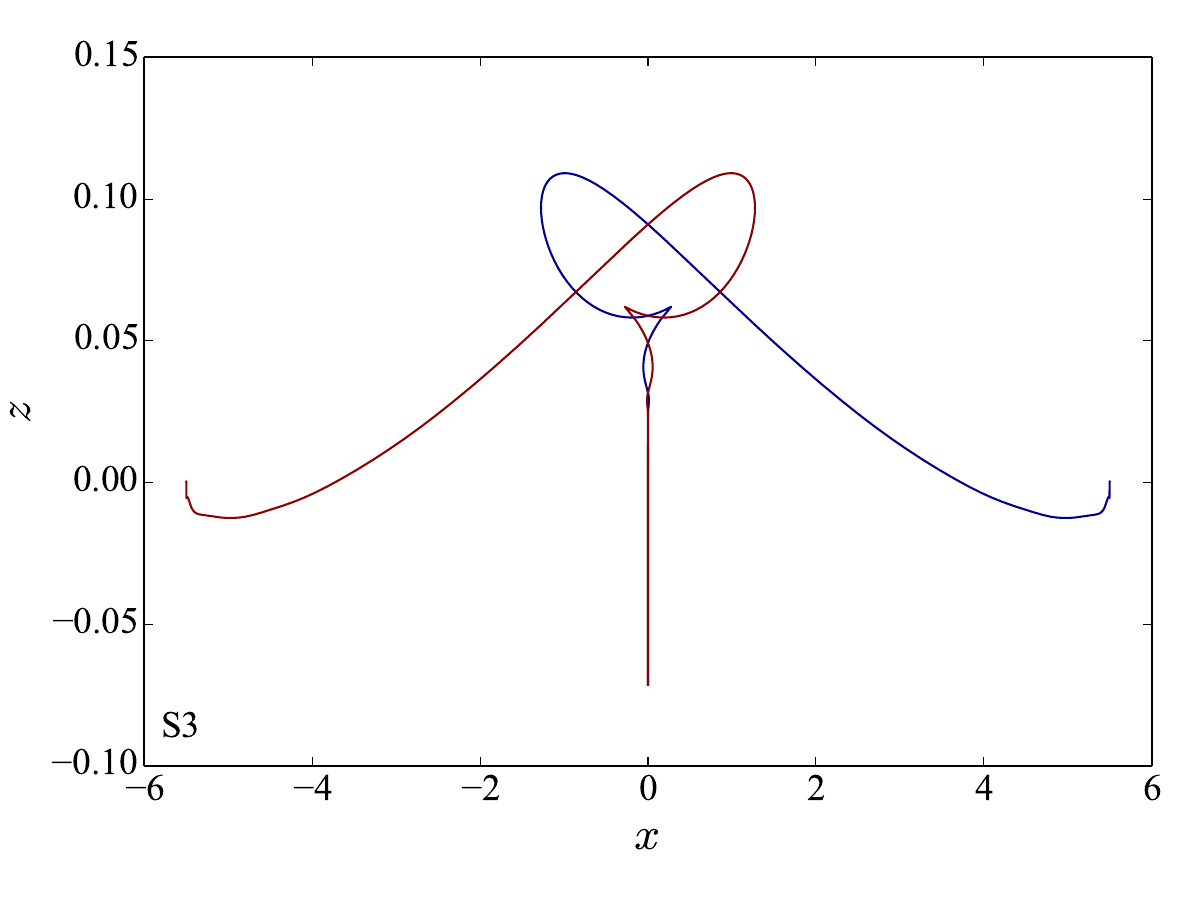}\\\includegraphics[width=0.329\linewidth]{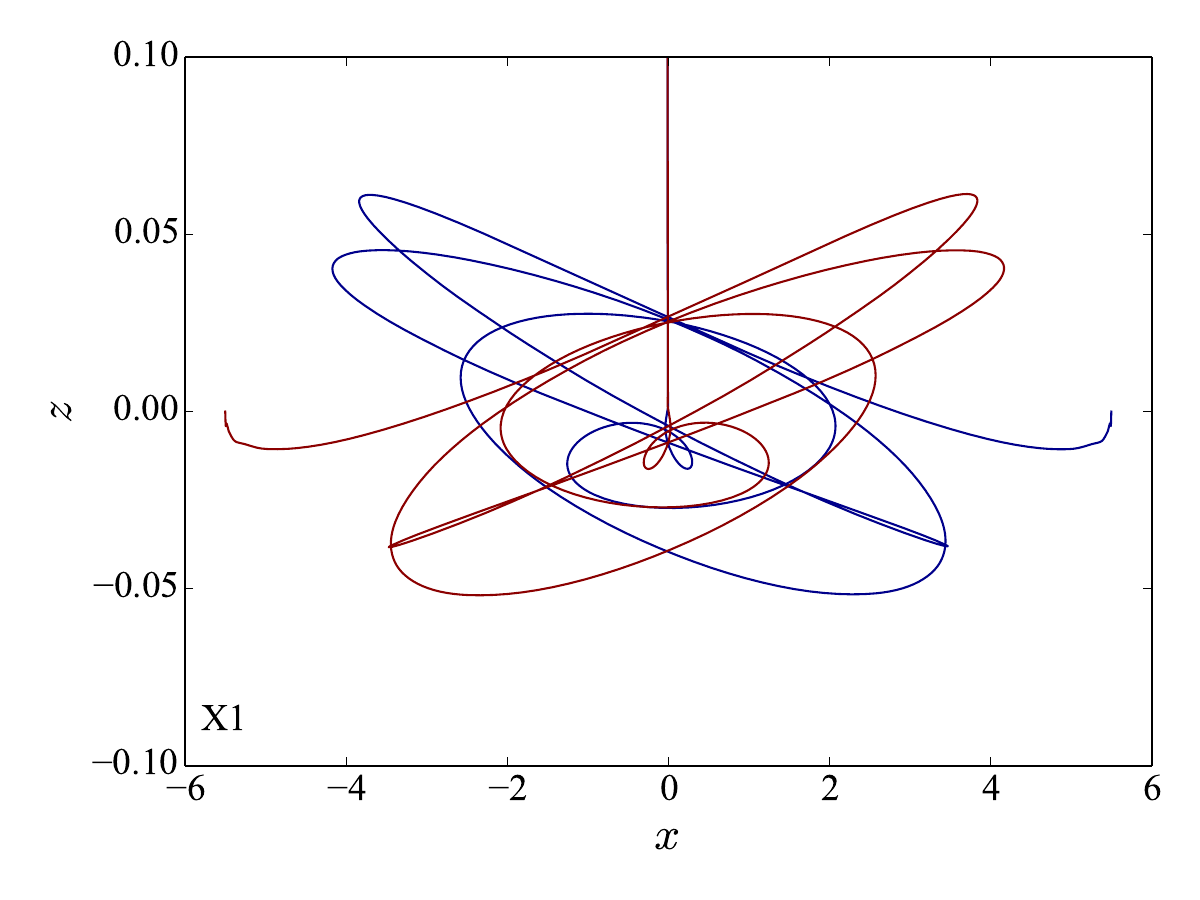}
\includegraphics[width=0.329\linewidth]{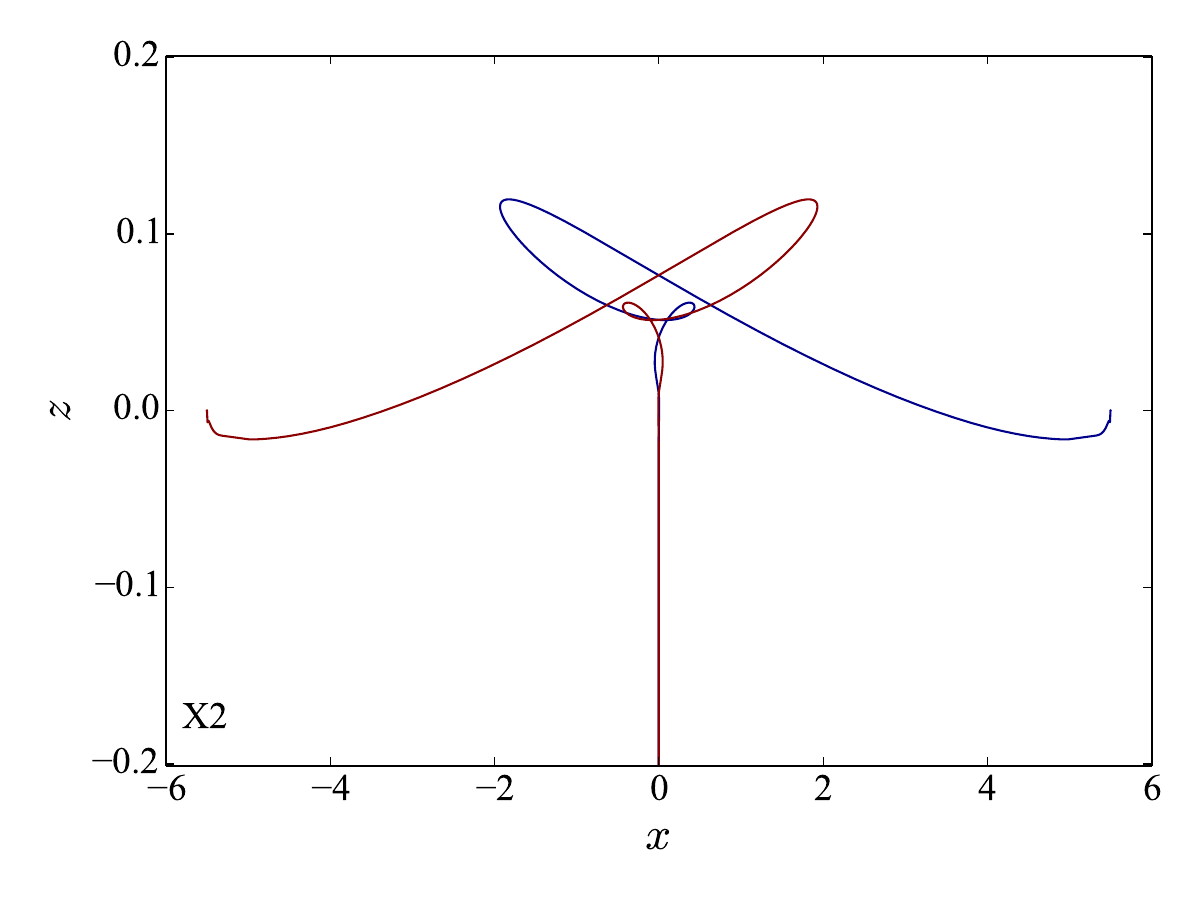}
\includegraphics[width=0.329\linewidth]{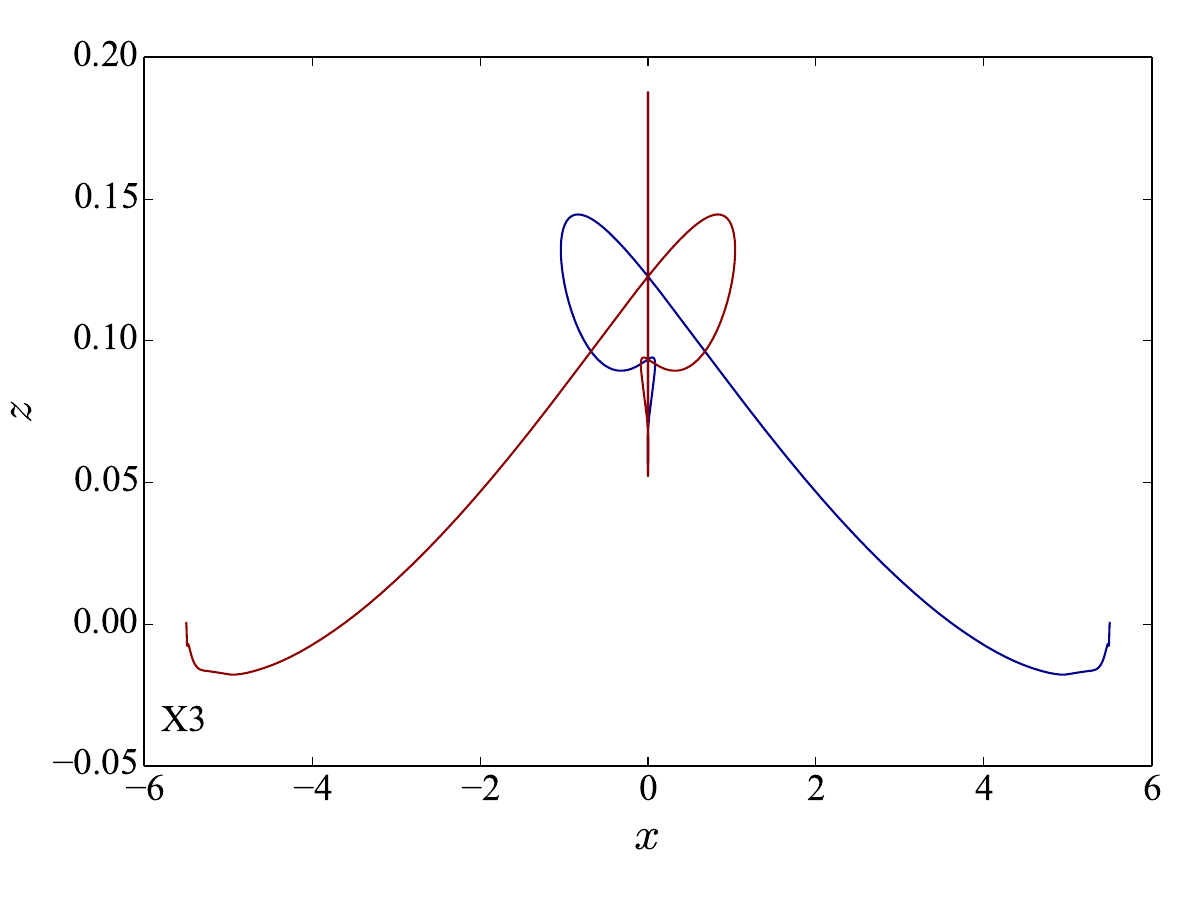}\\\includegraphics[width=0.329\linewidth]{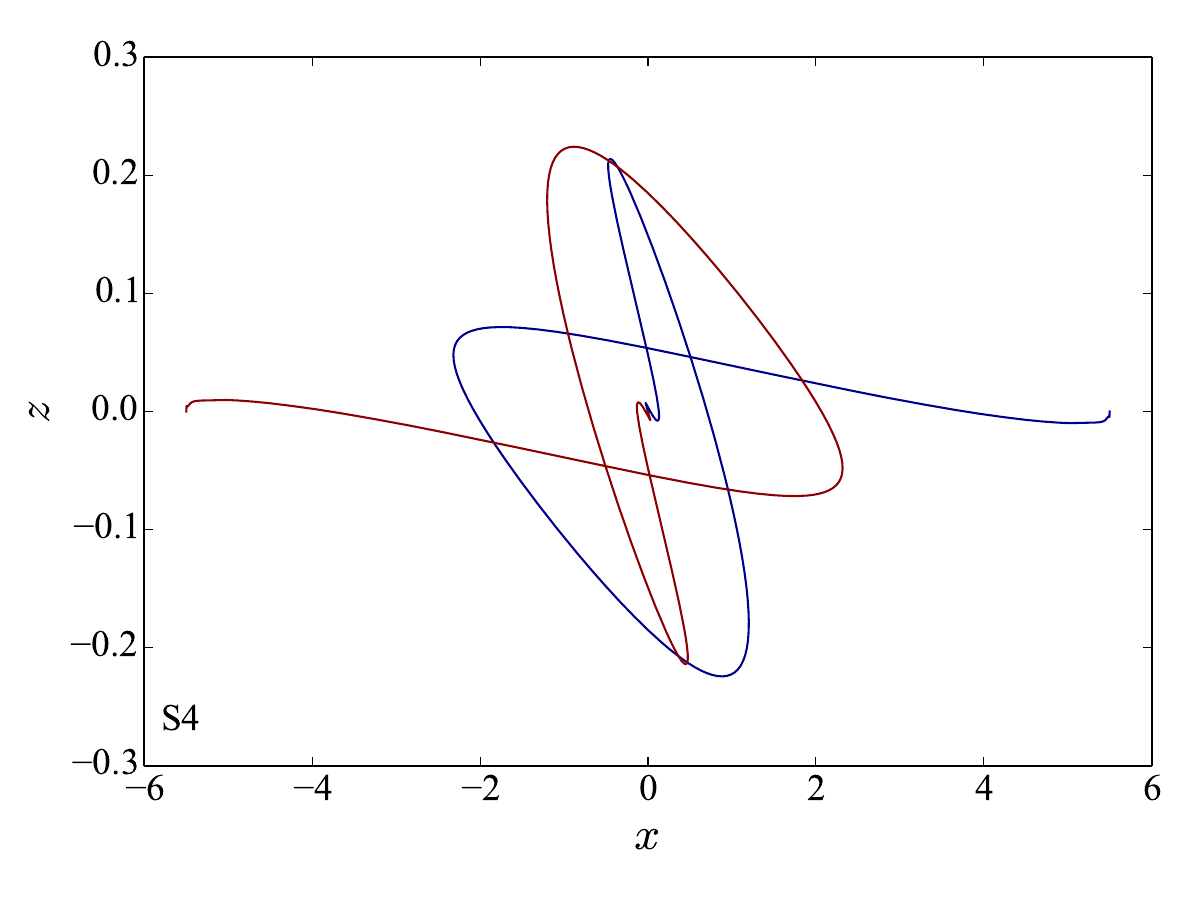}
\includegraphics[width=0.329\linewidth]{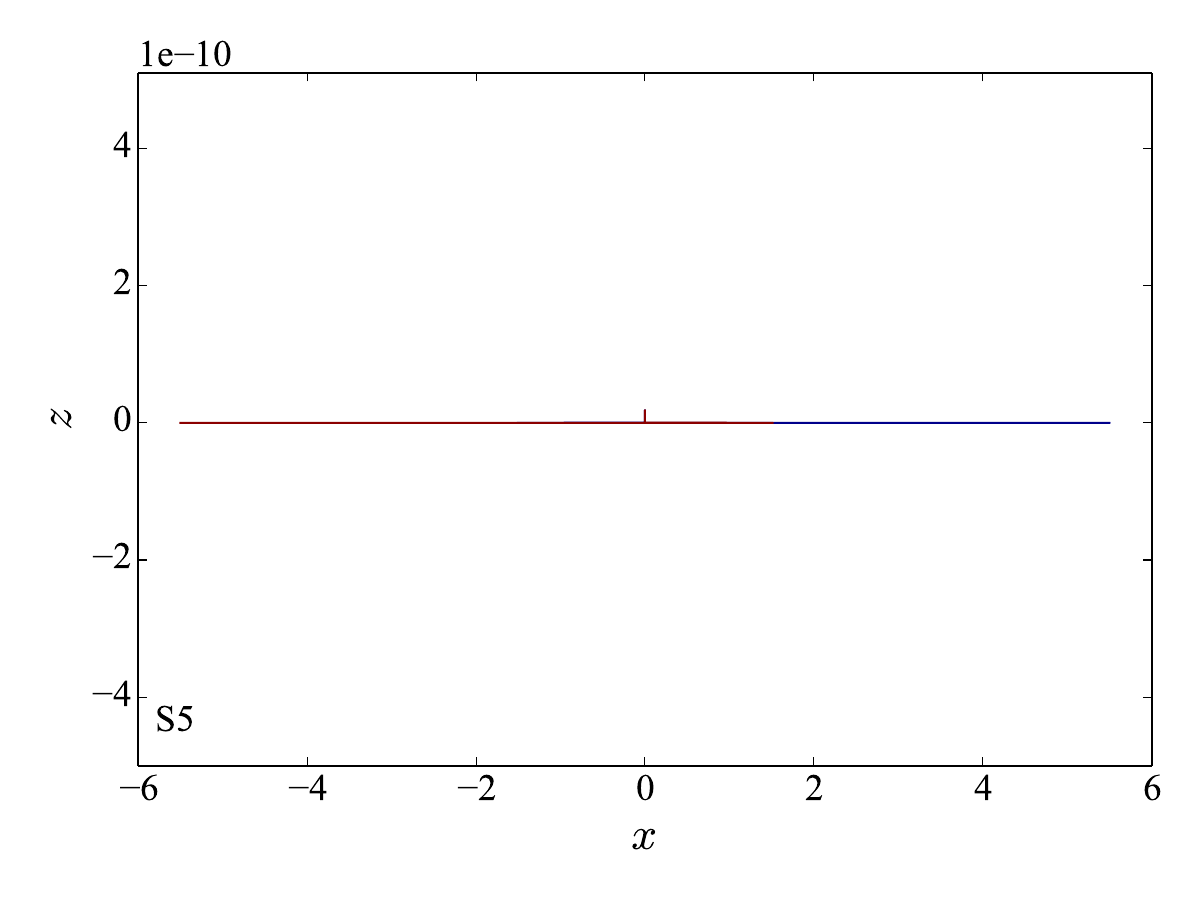}\includegraphics[width=0.329\linewidth]{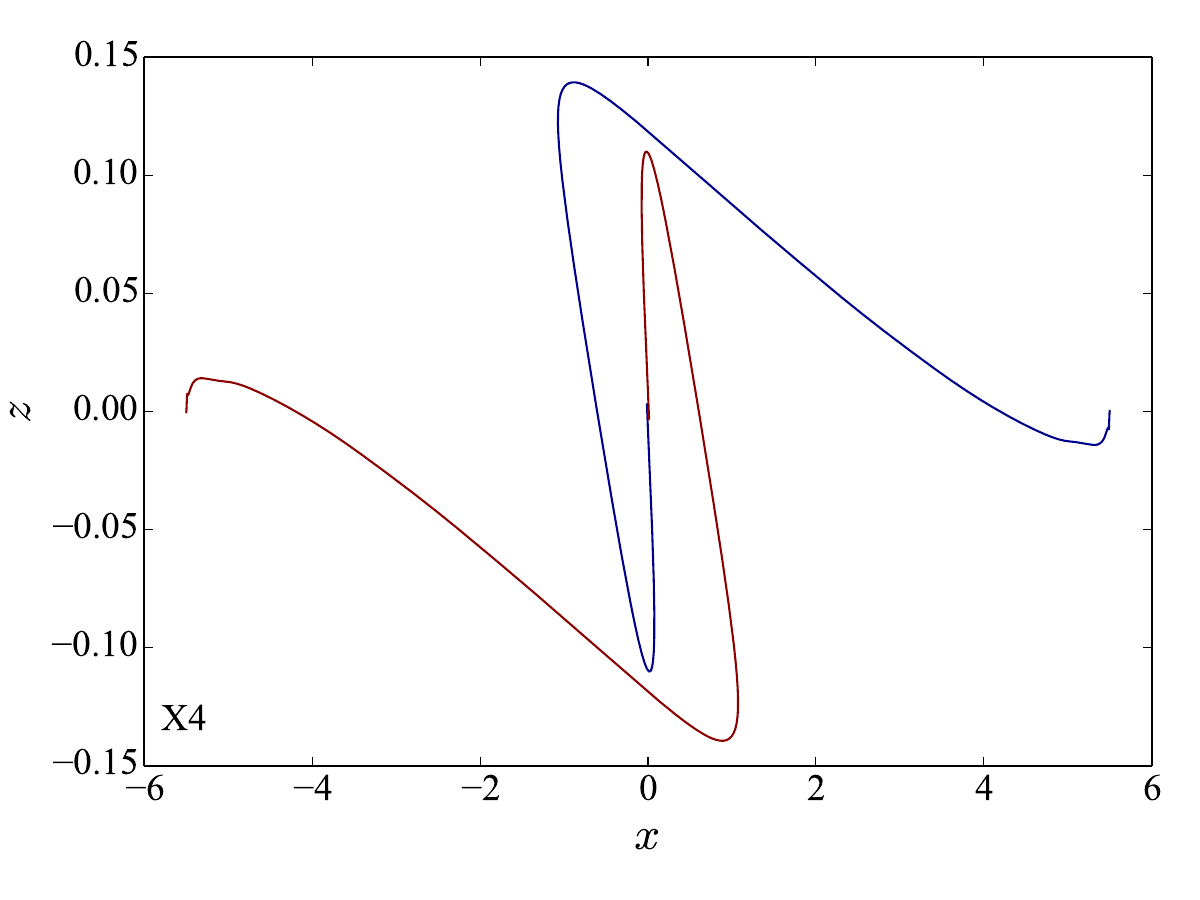}
\caption{Trajectories of the BHs {projected on} the $xz$ plane (with $y=0$) for the different configurations described in Table~\ref{tabhdbbh}. The existence of precession in the evolution is manifest in most cases. Configurations S1, S2, S3, X1, X2, and X3 give a non-zero contribution to the Chern-Porntryagin and also display a kick after the merger.}
\label{figTrajz}
\end{figure*}

\subsection{Orbital mergers}
While head-on collisions are useful to easily identify the role of the relative spin configurations  on the Chern-Pontryagin, as well as its connection to the lack of mirror symmetry in the problem, it is also interesting to study the more astrophysically relevant case of orbital binary BHs to take into account the contribution of the inspiral phase. As we will detail shortly, the main new feature in this case is the presence of oscillations {in the gravitational chirality} during the evolution.  
 
To explore this problem we perform  9 equal-mass and equal-spin magnitude binary Kerr BH mergers in eccentric orbits. As before, we set the initial separation at $D=11$, the initial momentum of the BHs to $|P_{x}|=0.000728$
and $|P_{y}|=0.0903$, and we vary their spin and mass. The details of each binary  configuration can be found in Table~\ref{tabhdbbh}. As a result of varying both spin and mass without changing the initial separation and initial linear momentum, the orbital dynamics are drastically modified and we obtain different eccentric motions for the binary. This is shown in Fig.~\ref{figTraj} where we plot the trajectories of the BHs in the equatorial plane ($z=0$). 
 
We have considered two main setups: in the first one, the spins are tilted 45 degrees with respect to the orbital plane and 90 degrees with respect to each other: $(\leftarrow,0,\uparrow),\, (\rightarrow,0,\uparrow)$ (configurations S1, S2, S3); while in the other the spins have only a $x$-component and are anti-aligned, $(\leftarrow,0,0),\, (\rightarrow,0,0)$, as in the head-on case (configurations X1, X2, X3). Three additional cases are considered for completeness: a binary with aligned spins but tilted 45 degrees with respect to the orbital plane $(\rightarrow,0,\uparrow),\, (\rightarrow,0,\uparrow)$ (configuration S4); a binary with both spins aligned with the orbital angular momentum $(0,0,\uparrow),\, (0,0,\uparrow)$ (configuration S5); and a binary with aligned spins in the $x$ direction  $(\rightarrow,0,0),\, (\rightarrow,0,0)$ (configuration X4).

\begin{figure}[t!]
\centering
\includegraphics[width=0.8\linewidth]{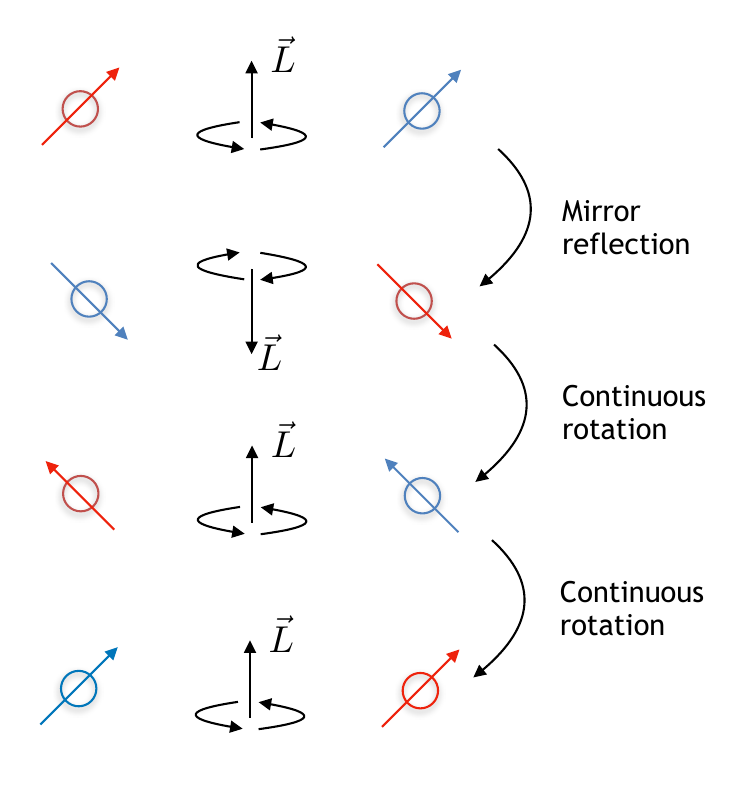}
\caption{Binary BH system with orbital evolution that may or may not yield a non-zero value of the Chern-Pontryagin (\ref{gravcp}). The picture represents one instant of time during the inspiral. The colors serve to distinguish the two BHs (each one with their own mass and spin magnitude). If the two BHs have the same masses and spins, then the system has a mirror symmetry which produces $F[g_{ab}](t)=\int_{\Sigma}d^3x\sqrt{-g}R_{abcd}{^*R}^{abcd}=0$ at the time considered. Numerical simulations confirm this theoretical prediction. However, if the two BHs are distinguishable, there is no mirror symmetry and $F[g_{ab}](t)\neq 0$.}
\label{45spin}
\end{figure}

Most of the configurations studied here precess because  the orientation of the spins have been chosen such that they are not aligned with the orbital angular momentum.  The effect of precession is shown in Fig.~\ref{figTrajz} where we plot the trajectories of the BHs in the $xz$ plane ($y=0$). In general, both BHs start at $z=0$ but move in the $z$ plane. The only exception is the S5 configuration which consists of two aligned Kerr BHs with the orbital plane (bottom row, middle panel of Figs.~\ref{figTraj} and~\ref{figTrajz}). Despite this, not all precessing evolutions give a non-zero value of the Chern-Pontryagin.  
For instance,  configurations S4 and X4 have the BH spins aligned with each other, but not aligned with the orbital angular momentum. In these cases, we also observe precession, which translates to non-negligible motion in the $z$ plane (bottom row of Fig.~\ref{figTrajz}). However, their contribution to the Chern-Pontryagin is compatible with zero  \footnote{{Note that a zero value cannot be strictly attained due to the numerical truncation errors. The non-zero values obtained for the S4, S5, and X4 configurations are the result of numerical accuracy, which, for the same resolution, may vary depending on the binary orbital dynamics and the precession they undergo. Our convergence study (see Appendix~\ref{appendix}) shows that such values do converge to zero as the resolution increases.}} (see Table~\ref{tabhdbbh}). This could have been anticipated using arguments of mirror symmetry (see Fig. \ref{45spin}). It should be noted though that the Chern-Pontryagin vanishes only because we are considering equal-mass equal-spin BHs in this problem. If the mass or spin magnitude were different, we would have a non-zero effect.

From all these observations we can conclude that if $\Delta \hat Q_5/\hbar\neq 0$ then the system is necessarily precessing, but the converse is not necessarily true. On the other hand, the configurations with misaligned spins move along the $z$-axis and suffer a {gravitational recoil or} kick after the merger \cite{gonzalez2007maximum,gonzalez2007supermassive} as seen in Fig.~\ref{figTrajz}. These configurations correspond precisely to the binaries that have a non-vanishing Chern-Pontryagin and would produce a flux of circularly polarized photons. Therefore, we find that there {seems to be a connection between a non-zero Chern-Pontryagin, precession, and kicks. Our results suggest that all configurations with a non-vanishing $\Delta \hat Q_5/\hbar$ precess and are prone to kicks. However, the opposite is not necessarily true. Note  that not all precessing systems give a non-zero contribution or suffer a kick, for instance the X4 configuration.}

\begin{figure*}%[t!]
\centering
\includegraphics[width=0.48\linewidth]{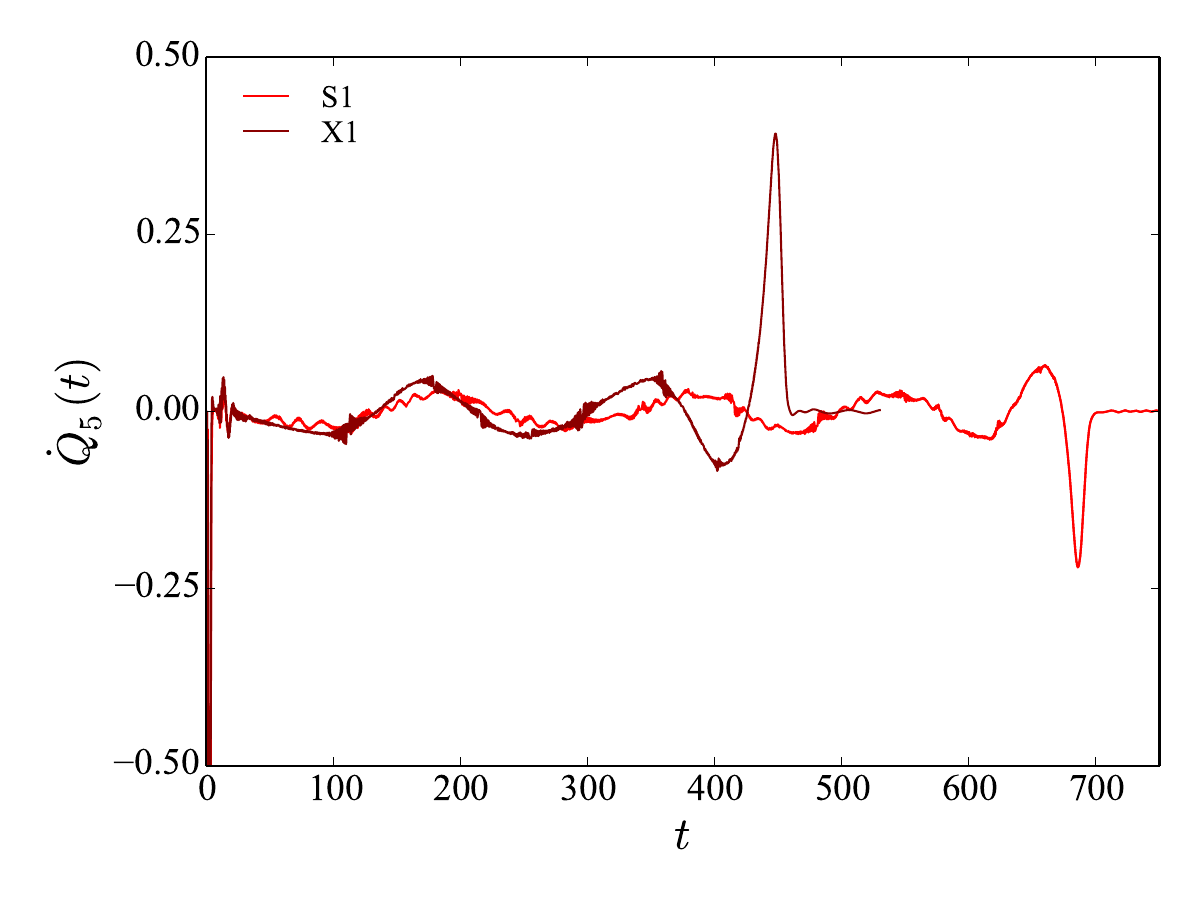}\includegraphics[width=0.48\linewidth]{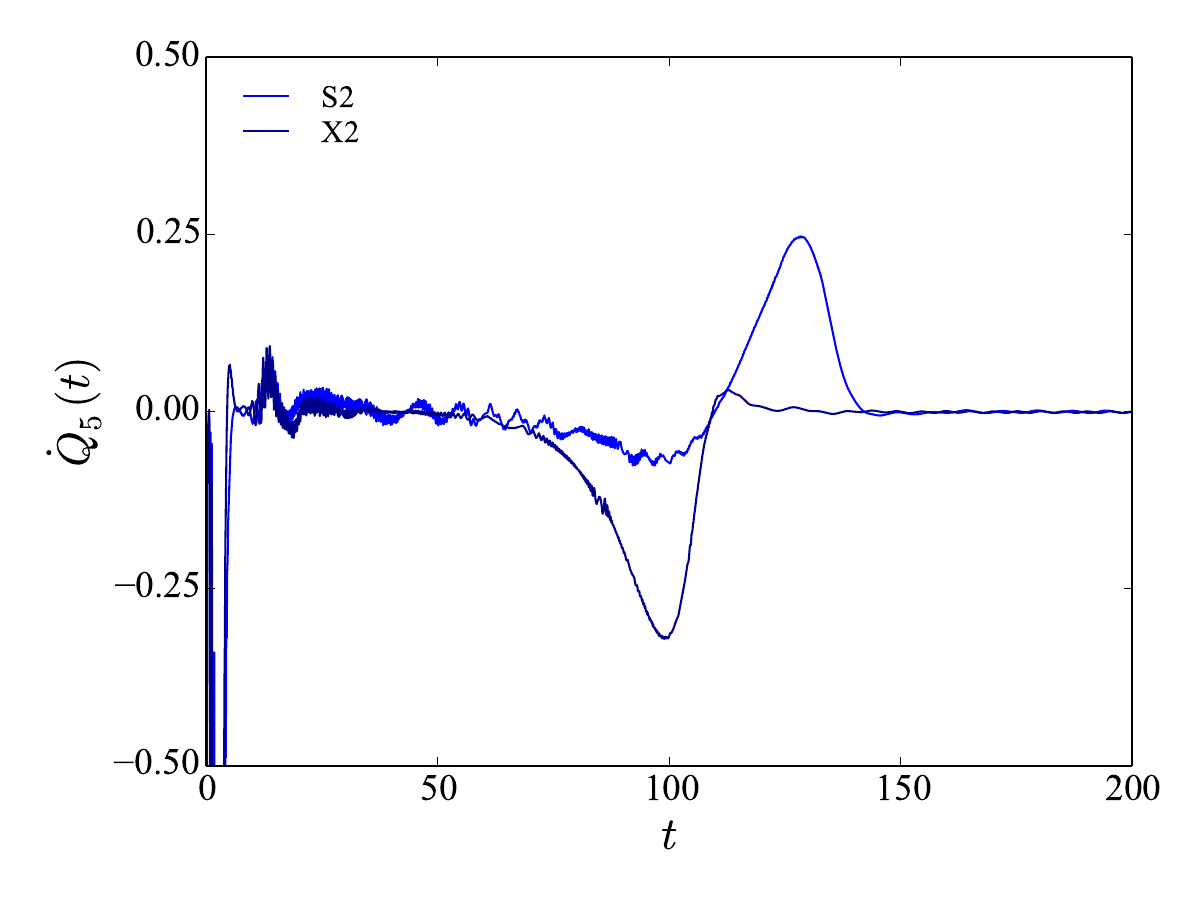}\\
\includegraphics[width=0.48\linewidth]{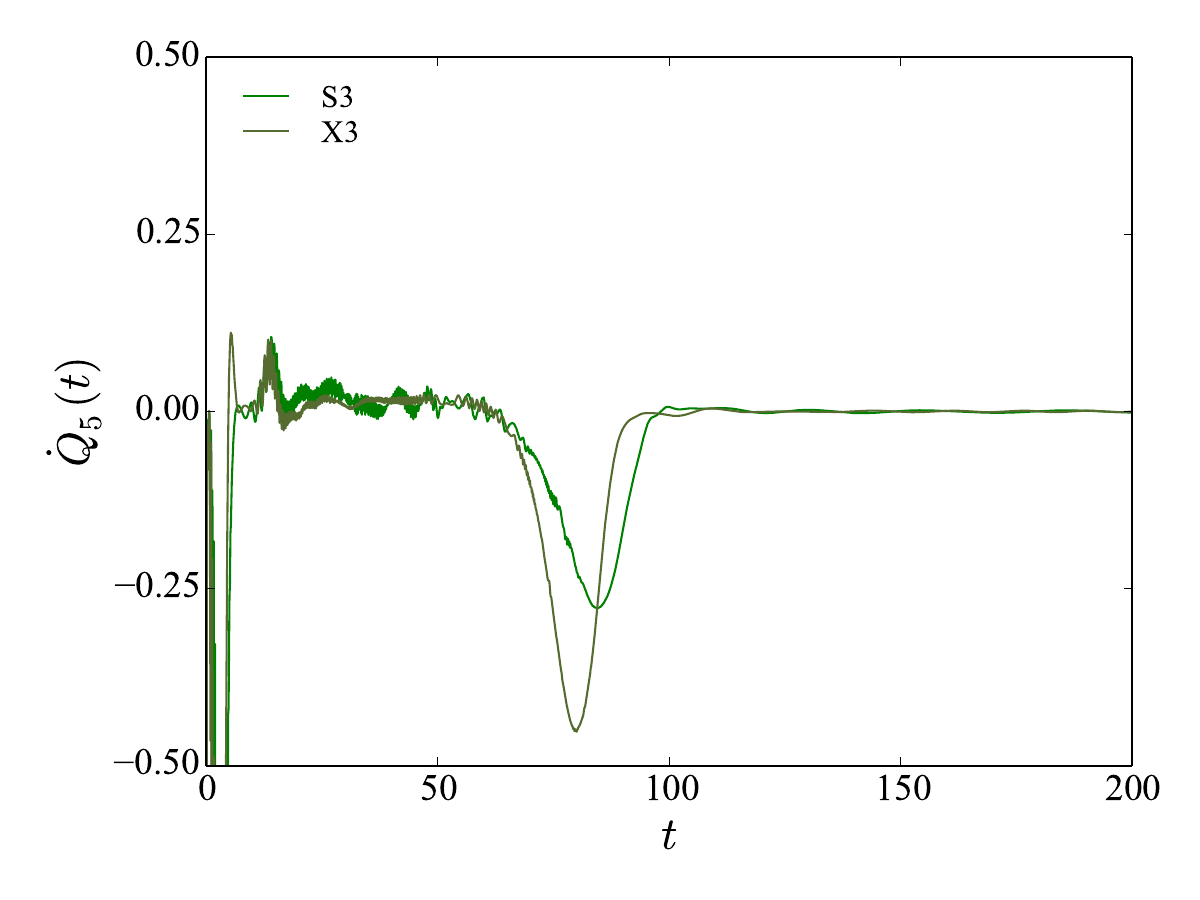}
\includegraphics[width=0.48\linewidth]{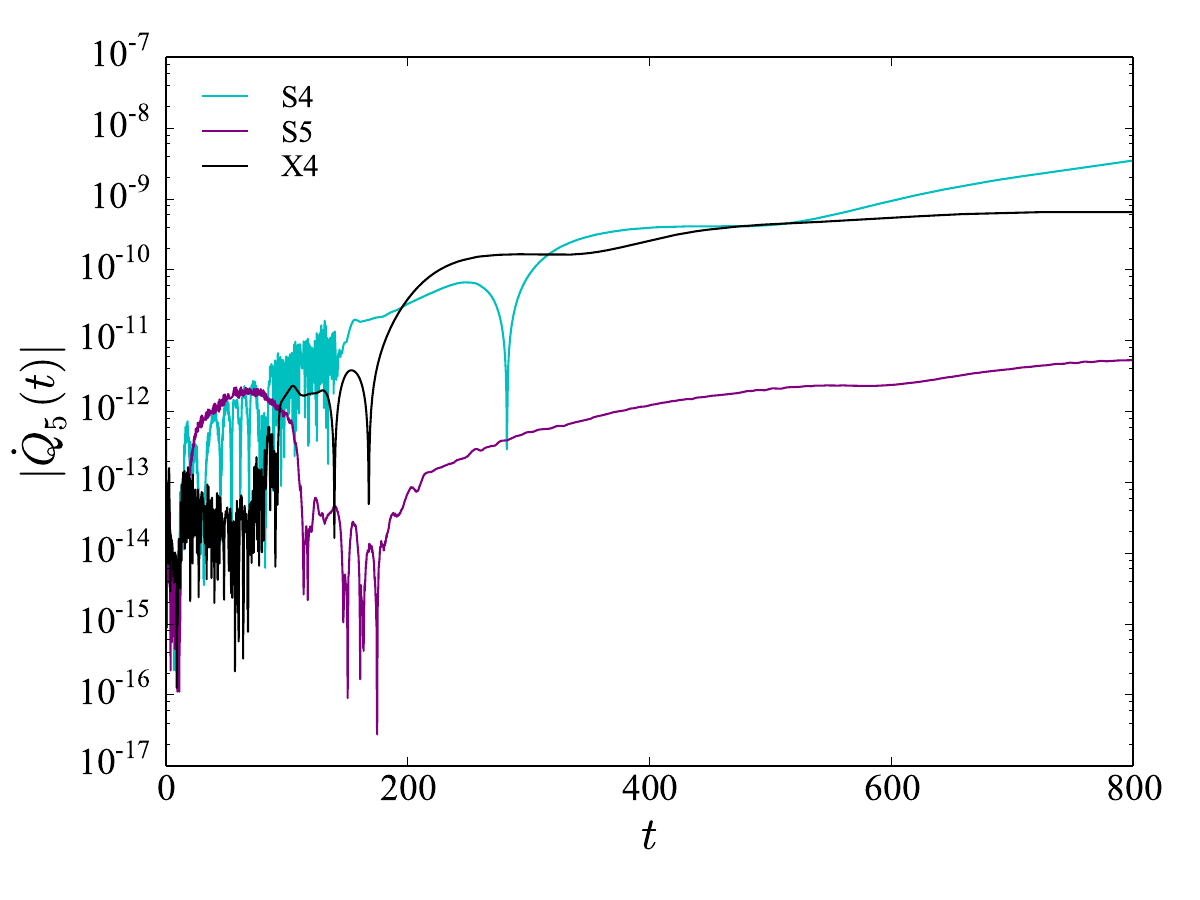}
\caption{Time evolution of $\dot Q_5$ computed from (\ref{Q5time}) for the nine binary BH configurations described in Table~\ref{tabhdbbh}. In cases when $\Delta\hat Q_{5}/\hbar\neq 0$ (top left and right panels, bottom left panel), $\dot Q_5(t)$ oscillates in time around zero during the inspiral phase all the way up to the merger. This periodicity in $\dot Q_5(t)$ is a manifestation of the cyclic evolution of the relative spin configuration of the two BHs during inspiral (see main text for details). The largest positive or negative peaks correspond to the time of merger, after which $\dot Q_5(t)$ drops down to zero, as expected for a stationary Kerr BH.}
\label{figEB}
\end{figure*}

\begin{figure}%[t!]
\centering
\includegraphics[width=1.0\linewidth]{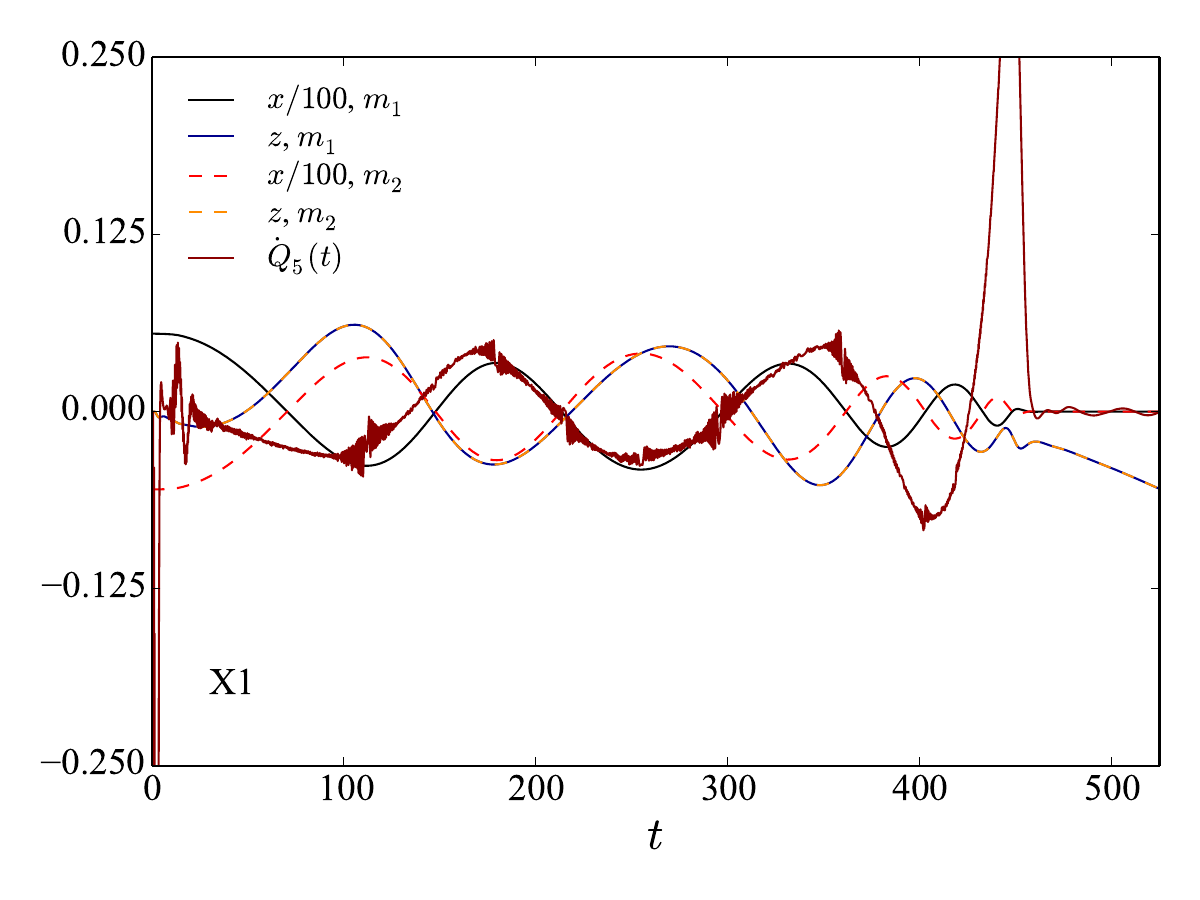}
\caption{Time evolution of $\dot Q_5$ computed from (\ref{Q5time}) for configuration X1 of Table~\ref{tabhdbbh} (solid dark red line). The $x$ (divided by 100) and $z$ coordinates of each  BH are also shown as a function of time. Notice the high correlation between the value of $\dot Q_5$ and the position of the BHs along the orbit. This supports the theoretical interpretation described in the main text.}
\label{figEBposed}
\end{figure}
\begin{figure}%[h!]
\centering
\includegraphics[width=1.0\linewidth]{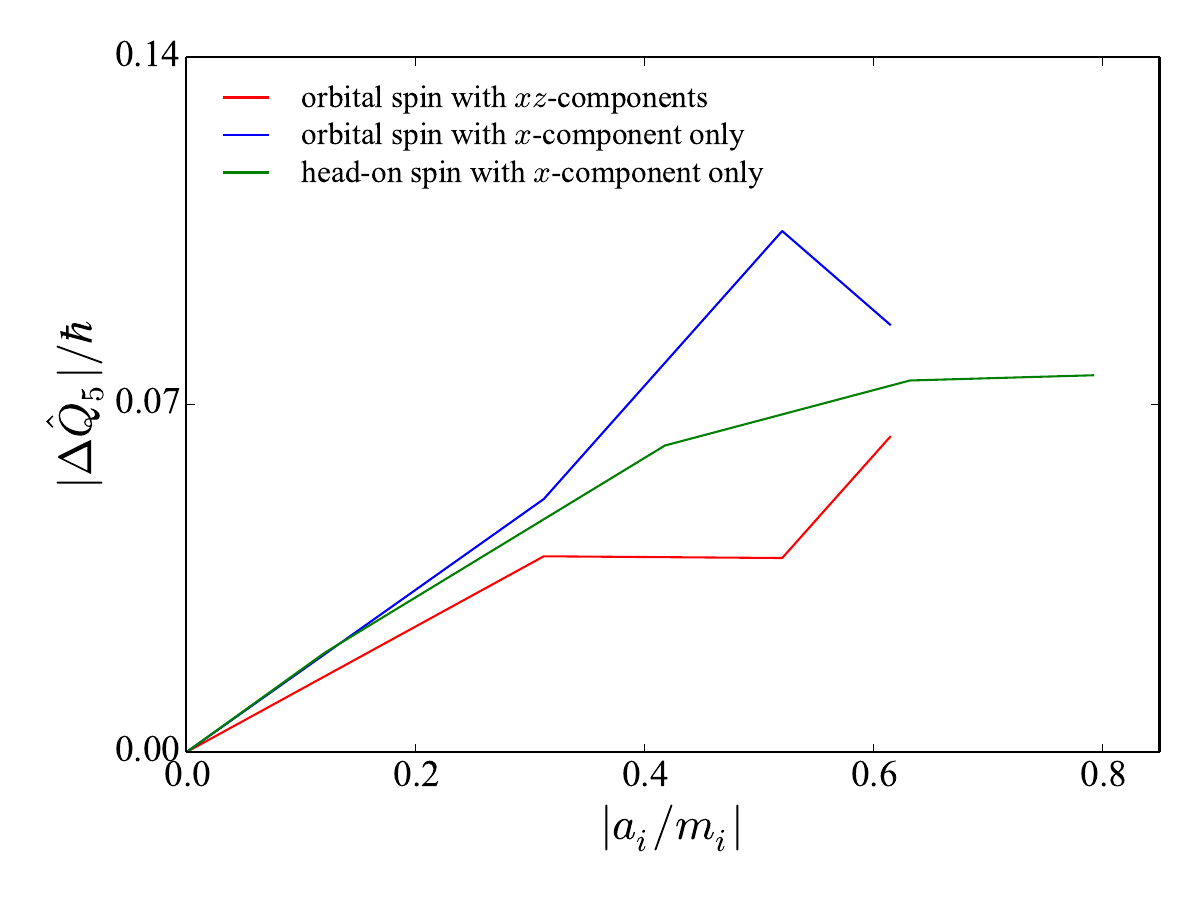}
\caption{Absolute value of $\Delta\hat Q_5/\hbar$ computed from (\ref{EB}) as function of the spin parameter $a_{i}/m_{i}$ for equal-mass and equal-spin orbital binary BH mergers with different spin inclinations with respect to the orbital plane: 45 degrees (red line) and 90 degrees (blue line). The green line corresponds to equal-mass and equal-spin head-on collisions with spins inclined 90 degrees.}
\label{figEBtotal}
\end{figure}
Since the Chern-Pontraygin (\ref{EB}) can be written as the time integral of a quantity that only depends on the geometry of the spatial slices in our 3+1 spacetime decomposition, it is interesting to see explicitly the evolution in time of the geometrical quantity
\begin{equation}
\dot Q_{5}(t) = \int_{\Sigma_t}d\Sigma_t\sqrt{-g}E_{ij}B^{ij}\, . \label{Q5time}
\end{equation}
{Notice that, in contrast to (\ref{gravcp}), this magnitude does depend on the total mass $M_*$ of the binary, as $1/M_*$. The integration in time cancels this dependence though.}

Figure~\ref{figEB}  shows the time evolution of this quantity for the 9 simulations. 
For instance, in the upper left panel, which corresponds to configurations S1 and X1 in Table~\ref{tabhdbbh}, we see that $\dot Q_{5}(t)$ oscillates around zero. This is in fact the general result in all orbiting cases where the Chern-Pontryagin is not zero. To understand correctly this behaviour we have to recall the analysis of symmetries of Sec.~II. Suppose a quasi-circular binary system. The relative orientation of the two BH spins evolve cyclically in time  during the inspiral. In particular, given a particular spin configuration at some instant of time $t_0$, with value $\dot Q_{5}(t_0)$, after half orbital period $T$ the new configuration of BH spins gets exactly the mirror-reflected version of the system at time $t_0$. If additionally the separation distance remains roughly constant during this half-period, we can expect $\dot Q_{5}(t_0+T/2)\approx -\dot Q_{5}(t_0)$.  Then, after one full orbital period, the relative spin configuration returns to the same state, and we can again expect $\dot Q_{5}(t_0+T)\approx \dot Q_{5}(t_0)$. The smaller the separation distance between the BHs, the greater this effect will be (because gravity gets more extreme). Therefore,  the oscillations are expected to increase adiabatically during the whole inspiral, until   {BHs merge} and we observe a sharp rise corresponding to the maximum peak in the plots. After the merger and formation of the final BH, the value goes to zero very quickly, as expected for a stationary Kerr BH. All these expectations are clearly {manifested} in Fig.~\ref{figEBposed}, which shows in more detail the time evolution of $\dot Q_5(t)$ together with the evolution of the $x$ and $z$ coordinates of both BHs. The maxima and minima of $\dot Q_5(t)$ are obtained approximately when the BHs are again located on the $x$ axis with $y=0$ (solid black and dashed red lines). {This is due to the initial setup, in which the initial spins are positioned in an extremal configuration for the Chern-Pontryagin ($\leftarrow,0,0$), ($\rightarrow,0,0$). Due to the orbital motion, when the black holes cross the $y$-axis (with $x=0$) the configuration becomes minimal ($\dot Q_5(t)$ vanishes) due to mirror symmetry. However, when they reach half an orbit and the black holes are back on the $x$-axis (with $y=0$), we find an extremal spin configuration but with opposite sign ($\rightarrow,0,0$), ($\leftarrow,0,0$). Finally, when the orbit is complete and again the black holes are located on the $x$-axis, we have the extremal configuration of the beginning (although the distance between the objects has been slightly reduced).} At similar times, their position in the z-axis $|z_i|$ (solid dark blue and orange dotted lines) also becomes maximum.  

The contribution of the inspiral  to the  Chern-Pontryagin $\Delta\hat Q_{5}/\hbar$ is small since, as commented above, the orbital motion changes the spin configuration cyclically and leads to consecutive positive and negative peaks in $\dot Q_5(t)$ that almost  cancel each other out after integrating in time. The most important contributions to $\Delta\hat Q_{5}/\hbar$ come from the last orbit and the merger. It is during the merger that we get the largest positive or negative peaks shown in the plots of Fig.~\ref{figEB}. In quasi-circular binaries (see upper left panel of Fig.~\ref{figEB}) there is also a previous large amplitude peak with opposite sign that can cancel an important part of the final maximum peak when computing the total time-integrated quantity. However, as the orbits become more and more eccentric (upper right and lower left panels of Fig.~\ref{figEB}), and in particular for head-on collisions, there is only one final peak. Therefore, it is for highly-eccentric collisions that the maximum net effect for $\Delta\hat Q_{5}/\hbar$ could be expected. On the other hand, the bottom right panel of Fig.~\ref{figEB} displays the time evolution of $\dot Q_{5}$ in cases when the two BH spins are aligned  (configurations S4, S5, X4), and for which the binary BH retains some mirror symmetry. In these cases not only the  Chern-Pontryagin $\Delta\hat Q_{5}/\hbar$ vanishes, as shown in Table~\ref{tabhdbbh}, but $\dot Q_5(t)$ is zero (within the numerical error) at all times, in excellent agreement with our theoretical interpretation in Sec.~II.

Finally, Fig.~\ref{figEBtotal} shows how the  Chern-Pontryagin (\ref{EB})  changes as a function of the spin parameter for different collisions and spin configurations. The comparison is not entirely accurate, since in the orbital case the trajectories are different for each binary and the final result may vary depending on the dynamics, but it serves as an illustrative estimate of the behaviour of the Chern-Pontryagin in these scenarios. The conclusion is that  (\ref{EB}) is maximized when the spins are as misaligned as possible with respect to each other and with respect to the orbital angular momentum. As expected, the more the mirror symmetry is broken in the binary, the higher the value of (\ref{EB}) is.

%%%%%%%%%%%%%%%%%%%%%%%%%%%%%
\section{Conclusions and further applications}
%%%%%%%%%%%%%%%%%%%%%%%%%%%%%

Motivated by quantum considerations, in this work we carried out a throughout  study of the Chern-Pontryagin curvature scalar (\ref{gravcp}) to figure out what are the key elements of binary BH systems that may trigger the spontaneous creation of photons with net helicity through quantum vacuum fluctuations. To this end we have performed a series of numerical simulations of head-one collisions and eccentric orbital mergers, with specific configurations of masses and spins motivated on arguments of mirror symmetry. Our findings indicate that orbital precession of the two BHs, or equivalently the misalignment of the two spins with the orbital angular momentum, can produce the required  helicity violation.

As remarked above, we solved the dynamical evolution of these BHs with numerical relativity simulations. However, the use of symmetry arguments has proven to be extremely efficient to understand correctly the numerical results. Our theoretical expectations have been validated one by one in the simulations. Given the highly non-linear nature of Einstein's equations and of the  systems involved, it is  remarkable that one can predict the outcomes of a quadratic curvature integral {over the whole spacetime} using simple arguments on mirror symmetry.
%As in particle physics and other branches of Physics
{In fact, the use of  symmetry breaking  may be helpful to gain further insights on the dynamics of binary BHs. More precisely}, the exploitation of mirror symmetry in Sec. II above allowed us to predict that a necessary condition for (\ref{gravcp}) and (\ref{GWcircular}) to be non zero is that the binary is precessing. Currently, the  identification of precessing binary BHs among all the observed events in LIGO-Virgo interferometers is an open problem and, although many events are expected to precess, there is only partial evidence of this in one single event GW200129 \cite{abbott2021gwtc,varma2022evidence,hannam2022general} (and in fact it is not free of controversy \cite{payne2022curious}). Precession is expected to produce a small modulation on the gravitational waveforms, but detecting this  requires more precision and  searches that include {this effect}~\cite{harry2016searching,bustillo2017detectability, PhysRevD.102.044035}.
Alternatively, symmetry arguments guarantee that if (\ref{GWcircular}) is not zero, then the binary is necessarily precessing. 
In other words, {the inference of {{\it net}}, non-negligible gravitational-wave circular polarization {from} LIGO-Virgo detections can be used  to identify} precessing systems
\footnote{{Notice that Eq. (\ref{GWcircular}) represents the {\it net}, circularly polarized flux of gravitational waves emitted by a binary, integrated among {\it all} directions on the sphere. While   non-precessing binaries can generate a gravitational-wave mode ($\ell, m$) with circular polarization, i.e. $| h_+^{\ell m}(\omega)-ih^{\ell m}_{\times}(\omega)|^2
-|h_+^{\ell m}(\omega) + ih^{\ell m}_{\times}(\omega)|^2\neq 0$, the mirror-symmetric mode ($\ell, -m$) cancels this contribution upon summation in (\ref{GWcircular}). An unbalance is only obtained when the binary black hole is precessing, i.e. when the mirror symmetry is broken. }}. 
This independent observable may pave the way for identifying precession sistematically. We plan to explore this possibility in future works~\cite{bustillo2023}.

Another interesting feature is that there seems to be a correlation between precessing binaries with non-zero Chern-Pontryagin and kicks due to gravitational-wave emission. 
{This is somewhat expected:}
on the one hand, kicks can be measured from the gravitational waves emitted by the system~\cite{bustillo2017detectability} and are expected to be originated from an asymmetry in the direction of the gravitational emission, that pushes the BH out of the orbital plane due the gravitational waves carrying linear momentum~\cite{tichy2007binary}. On the other hand, if the positive and negative modes $m$ of the spin-weighted spherical harmonics do not compensate each other, this is a indication of mirror asymmetry and therefore the Chern-Pontryagin~\cite{dRSGMAFNS} is different from zero. It may be possible that in some cases both asymmetries are connected~\cite{PhysRevLett.121.191102,bustillo2022gw190412}.

From a quantitative point of view, the results obtained for helicity violation in photons are rather small, the order of magnitude is similar to the Hawking radiation effect, roughly one photon of difference between the right-handed and left-handed fluxes for each merger. This is not really surprising, and taken at face value, it seems very unlikely that one may be able to observe this quantum effect directly for one single event. However, it should be noted that small numbers can seed macroscopic effects through classical amplification mechanisms. Besides, in large enough numbers the quantum effect may lead to significant implications. More precisely, if the formation channels of binary black holes in astrophysics favour  ``right-handed'' spin configurations over ``left-handed'', or viceversa, this may produce an accumulated effect in the universe. This is out of the scope of the present paper, our plan is to investigate this in more detail in the future \cite{bustillo2023}.\bigskip

\section{Acknowledgements}
We thank José A. Font, Ivan Agullo and José Navarro-Salas for useful discussions, and specially Juan Calderón Bustillo for  discussions regarding the current experimental evidence of precession in binary mergers. NSG is supported by the Spanish Ministerio de Universidades, through a María Zambrano grant (ZA21-031) with reference UP2021-044, funded within the European Union-Next Generation EU. ADR is supported through a M. Zambrano grant (ZA21-048) with reference UP2021-044 from the Spanish Ministerio de Universidades, funded within the European Union-Next Generation EU. This work has further been supported by  the European Horizon Europe staff
exchange (SE) programme HORIZON-MSCA-2021-SE-01 Grant No.~NewFunFiCO-101086251. This work is also supported by the Spanish Agencia Estatal de Investigaci\'on (Grant PID2021-125485NB-C21), by the Spanish Grant PID2020-116567GB-C21 funded by MCIN/AEI/10.13039/501100011033 and the project PROMETEO/2020/079 (Generalitat Valenciana). NSG thankfully acknowledges the computer resources at Tirant and the technical support provided by UV (RES-FI-2022-3-0006).

\begin{appendix}
\section{Code assessment}
\label{appendix}

\begin{figure}[t!]
\centering
\includegraphics[height=2.45in]{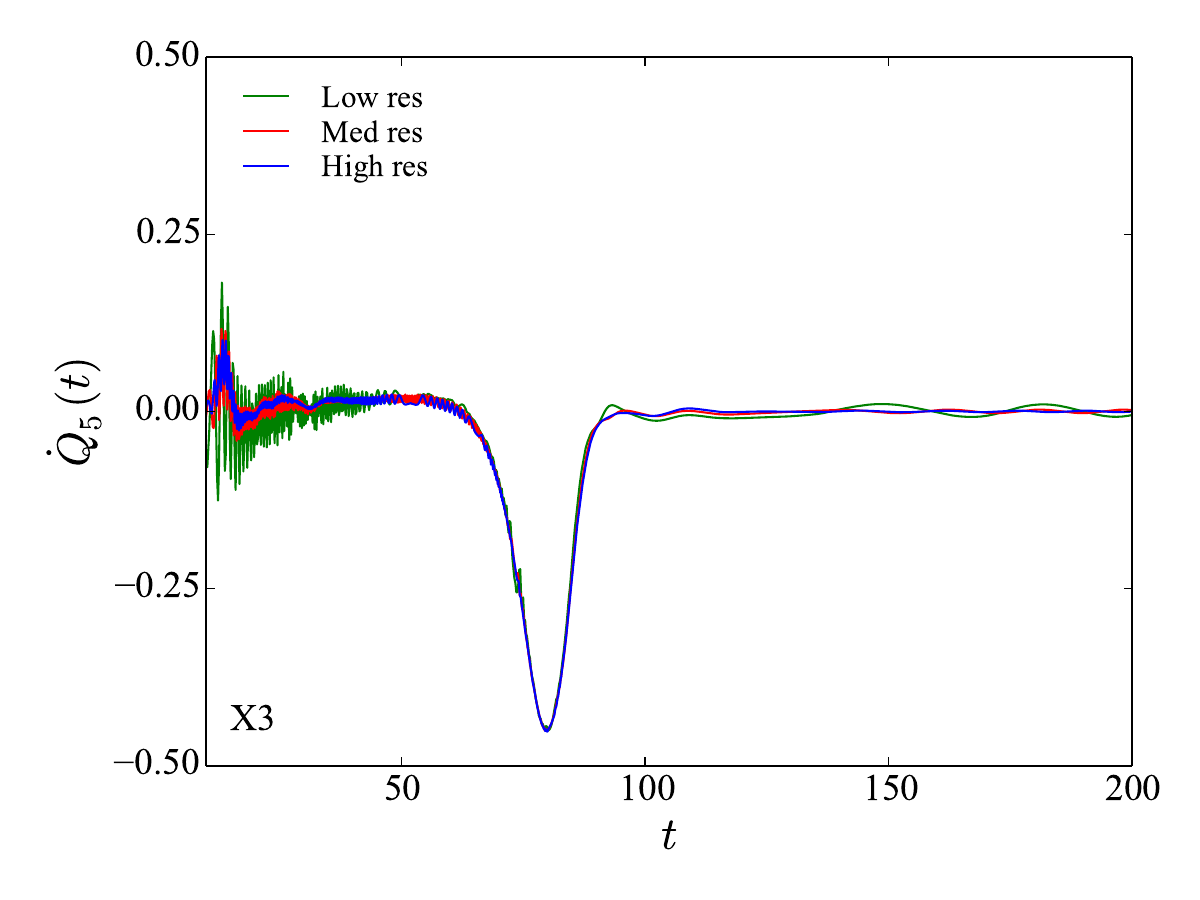}
\includegraphics[height=2.45in]{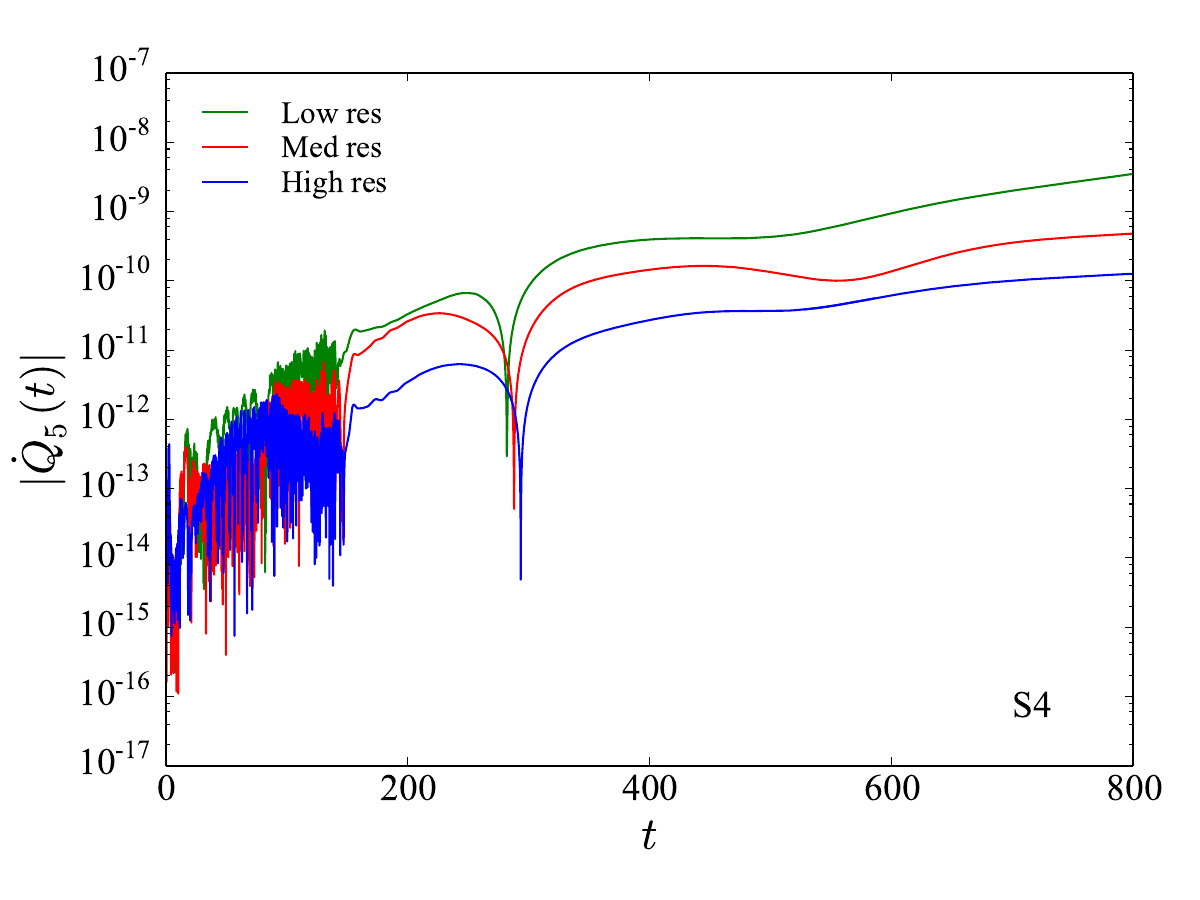}
\caption{Time evolution of $\dot Q_5$ for configurations X3 (top panel) and S4 (bottom panel) using three different resolutions with $dx=dy=dz=\lbrace0.03125,0.01953125,0.015625\rbrace$ in the finest level.}
\label{fig:convergence}
\end{figure}

{We briefly comment here on the convergence analysis we carried out to assess the quality of our simulations. To perform the binary black hole evolutions we have employed the freely-available {\tt Einstein Toolkit} code. Further convergence tests can be found in~\cite{toolkit2012open,loffler2012einstein}. In Fig.~\ref{fig:convergence} we plot the volume integral of the Chern-Pontryagin as a function of time $\dot Q_5(t)$ computed for configurations S4 and X3, which correspond to precessing systems with aligned and non-aligned spins respectively, using three different resolutions with $dx=dy=dz=\lbrace0.03125,0.01953125,0.015625\rbrace$ in the finest level. In the bottom panel of Fig.~\ref{fig:convergence} we show that the Chern-Pontryagin converges to zero at the expected fourth-order rate for the S4 configuration, confirming our symmetry analysis.}

\end{appendix}

\bibliographystyle{utphys}
\bibliography{References}

\end{document}